\documentclass{aa}
\usepackage{epsfig}
\begin{document}
\newcommand{\bea}{\begin{eqnarray}}    
\newcommand{\eea}{\end{eqnarray}}      
\newcommand{\be}{\begin{equation}}
\newcommand{\ee}{\end{equation}}
\newcommand{\bef}{\begin{figue}}
\newcommand{\eef}{\end{figure}}
\newcommand{\etal}{et al.}
\newcommand{\kms}{\,{\rm km}\;{\rm s}^{-1}}
\newcommand{\hubunits}{\,\kms\;{\rm Mpc}^{-1}}
\newcommand{\hmpc}{\,h^{-1}\;{\rm Mpc}}
\newcommand{\hkpc}{\,h^{-1}\;{\rm kpc}}
\newcommand{\msun}{M_\odot}
\newcommand{\K}{\,{\rm K}}
\newcommand{\cm}{{\rm cm}}
\newcommand{\cd}{{\langle n(r) \rangle_p}}
\newcommand{\Mpc}{{\rm Mpc}}
\newcommand{\kpc}{{\rm kpc}}
\newcommand{\xir}{{\xi(r)}}
\newcommand{\xrp}{{\xi(r_p,\pi)}}
\newcommand{\xsirpi}{{\xi(r_p,\pi)}}
\newcommand{\wrp}{{w_p(r_p)}}
\newcommand{\gr}{{g-r}}
\newcommand{\Navg}{N_{\rm avg}}
\newcommand{\Mmin}{M_{\rm min}}
\newcommand{\fiso}{f_{\rm iso}}
\newcommand{\Mr}{M_r}
\newcommand{\rp}{r_p}
\newcommand{\zmax}{z_{\rm max}}
\newcommand{\zmin}{z_{\rm min}}

\def\eg{{e.g.}}
\def\ie{{i.e.}}
\def\spose#1{\hbox to 0pt{#1\hss}}
\def\ltapprox{\mathrel{\spose{\lower 3pt\hbox{$\mathchar"218$}}
\raise 2.0pt\hbox{$\mathchar"13C$}}}
\def\gtapprox{\mathrel{\spose{\lower 3pt\hbox{$\mathchar"218$}}
\raise 2.0pt\hbox{$\mathchar"13E$}}}
\def\inapprox{\mathrel{\spose{\lower 3pt\hbox{$\mathchar"218$}}
\raise 2.0pt\hbox{$\mathchar"232$}}}

\title{Power law correlations in galaxy distribution 
and finite volume effects from the Sloan Digital Sky Survey Data Release
Four}

\subtitle{}

\author{Francesco
 Sylos Labini \inst{1,2},   Nickolay  L. Vasilyev \inst{3} and Yurij V. Baryshev \inst{3}
}

\titlerunning{Galaxy correlations from the DR4 of SDSS}
\authorrunning{Sylos Labini, Vasilyev, Baryshev}

\institute{``Enrico Fermi Center'', Via Panisperna 89 A, 
Compendio del Viminale, 00184 Rome, Italy
\and``Istituto dei Sistemi Complessi'' CNR, 
Via dei Taurini 19, 00185 Rome, Italy 
\and 
Institute of Astronomy, St.Petersburg 
State University, Staryj Peterhoff, 198504, St.Petersburg, Russia}

\date{Received / Accepted}

\abstract{
We discuss the estimation of galaxy correlation properties in several
volume limited samples, in different sky regions, obtained from the
Fourth Data Release of the Sloan Digital Sky Survey.  The small scale
properties are characterized through the determination of the nearest
neighbor probability distribution. By using a very conservative
statistical analysis, in the range of scales [0.5,$\sim$ 30] Mpc/h we
detect power-law correlations in the conditional density in redshift
space, with an exponent $\gamma=1.0 \pm 0.1$. This behavior is stable
in all different samples we considered thus it does not depend on
galaxy luminosity. In the range of scales [$\sim$ 30,$\sim$ 100] Mpc/h
we find evidences for systematic unaveraged fluctuations and we
discuss in detail the problems induced by finite volume effects on the
determination of the conditional density. We conclude that in such
range of scales there is an evidence for a smaller power-law index of
the conditional density. However we cannot distinguish between two
possibilities: (i) that a crossover to homogeneity (corresponding to
$\gamma=0$ in the conditional density) occurs before 100 Mpc/h, (ii)
that correlations extend to scales of order 100 Mpc/h (with a smaller
exponent $0 < \gamma <1$). We emphasize that galaxy distributions in
these samples present large fluctuations at the largest scales probed,
corresponding to the presence of large scale structures extending up
to the boundaries of the present survey. Finally we discuss several
differences between the behavior of the conditional density in mock
galaxy catalogs built from cosmological N-body simulations and real
data. We discuss some theoretical implications of such a fact
considering also the super-homogeneous features of primordial density
fields.
}
\maketitle


\section{Introduction}

One of the main open problems in modern cosmology is represented by
the statistical characterization and the physical understanding of
large scale galaxy structures.  The first question in this context
concerns the studies of galaxy correlation properties.  Particularly
two-point properties are useful to determine  correlations
and their spatial extension.  There are different ways of measuring
two-point properties and, in general, the most suitable method depends
on the type of correlations, strong or weak, characterizing a given
point distribution in a certain sample.

For example, Hogg et al. (2005) have recently measured the conditional
average density in a sample of Luminous Red Galaxies (LRG) from a
data release of the Sloan Digital Sky Survey (SDSS).  Such a
statistics is very useful to determine correlation properties in the
regime of strong clustering and the spatial extension of strong
fluctuations in a given sample. This was firstly introduced by
Pietronero (1987) and then measured in many samples by Sylos Labini et
al. (1998). We refer the reader to Baryshev \& Teerikorpi (2005) for a
review of the measurements of the reduced and complete
correlation functions by different authors in the various angular and
three-dimensional samples.

The conditional density gives the average density of points in a
spherical volume (or a spherical shell) centered around a galaxy (see
Gabrielli et al. 2004 for a discussion about this method).  The
results obtained by Hogg et al. (2005) can be summarized as follows:

(i) A simple power-law scaling corresponding to a correlation exponent
$\gamma \approx 1$ gives a very good fit to the data up to at least
$20$ Mpc/h, over approximately a decade in scale.  We note 
that these results are in good agreement with those obtained
by Sylos Labini et al.  (1998) through the analyses of many smaller 
samples and more recently by Vasilyev, Baryshev and Sylos Labini
(2006) in the 2dFGRS.
 
(ii) The second important result of Hogg et al. (2005) is that at
larger scales (i.e. $r >30$ Mpc/h) the conditional density continues
to decrease, but less rapidly, until about $\sim 70$ Mpc/h, above
which it seems to flatten up to the largest scale probed by the sample
($100$ Mpc/h).  The transition between the two regimes is slow, in the
sense that the conditional density at $\sim 20$ Mpc/h is about twice
the asymptotic mean density. Joyce et al. (2005) have discussed the
basic implications of these results noticing, for example, that the
possible convergence to a well defined homogeneity in a volume
equivalent to that of a sphere of radius 70 Mpc/h, place in doubt
previous detections of ``luminosity bias'' from measures of the
amplitude of the reduced correlation function $\xi(r)$.  They
emphasized that the way to resolve these issues is to first use, in
volume limited (VL) samples corresponding to different ranges of
luminosity, the conditional density to establish the features of
galaxy space correlations. Note that Sylos Labini et al. (1998) found
evidences for a continuation of the small scale power-law to distances
of order hundreds of Mpc/h, although with a weaker statistics, which
seems to be not confirmed by Hogg et al. (2005).

In this paper we continue the analysis of galaxy distributions
previously applied to the 2dFGRS data (Vasilyev et al. 2006) to the
so-called ``main galaxy sample'' of SDSS Data Release (DR4), in
the spirit of the tests discussed above.  In a companion paper we will
discuss the properties of the LRG sample of the SDSS DR4, which can be
directly compared with the results of Hogg et al. (2005) and
Eiseinstein et al. (2005).

The paper is organized as follows. In Section 2 we describe the data
and the way we have constructed the VL samples. We also discuss the
determination of the nearest neighbor (NN) distribution, and of the
average distance between nearest galaxies, which allows us to define
the lower cut-off for the studies of correlations. In addition we
discuss the determination of the radial counts in different VL
samples, emphasizing that large variations for this quantity are found
in the different samples.  Such fluctuations, which seem to be
persistent up to the sample boundaries, correspond to the large scales
structures observed in these catalogs. The quantitative
characterization of the correlation properties of these fluctuations
is presented in Section 3, where we discuss the determination of the
conditional average density in the different VL samples. In particular
we present several tests useful to clarify the effect of systematic
fluctuations at scales of order of the samples size.

In Section 4 we discuss the differences between the galaxy conditional
density, measured in these samples and the conditional density of
point-particles in cosmological N-body simulations. We show that by
using this statistics, together with a study of the NN probability
distribution, two-point properties of observed galaxies of different
luminosity and mock galaxy catalogs constructed using particles
lying in region with different local density in cosmological N-body
simulations, present different behaviors.  Finally in Section 5 we
draw our main conclusions.


\section{The data} 

The SDSS (http://www.sdss.org) is currently the largest spectroscopic
survey of the extragalactic objects and one of the most ambitious
observational programs ever undertaken in astronomy. It will measure
about 1 million redshifts, giving a complete mapping of the local
universe up to a depth of several hundreds of Mpc.  In this paper we
consider the data from the latest public data release (SDSS DR4) which
is accessible at http://www.sdss.org/dr4 (Adelman-McCarthy et
al. 2005) containing redshifts for more than 565 thousands of galaxies
and 67 thousands of quasars.  There are two independent parts of the
galaxy survey in the SDSS: the main galaxy sample and the LRG
sample. Here we discuss the former only.  The spectroscopic survey
covers an area of 4783 square degrees of the celestial sphere. The
apparent magnitude limit for the galaxies is 17.77 in the $r$-filter
and photometry for each galaxy is available in five different bands,
of which we consider the ones in the $r$ and $g$ filters.

\subsection{Definition of the samples}

We have used the following criteria to query the SDSS DR4
database. First of all we constrain the flags indicating the type of
object so that we select only the objects from the main galaxy sample.
We then consider galaxies in the redshift interval $10^{-4} \leq z
\leq 0.3$ and with the redshift confidence parameter larger than
$0.95$.  In addition we apply the filtering condition $r < 17.77$,
thus taking into account the target magnitude limit for the main
galaxy sample in the SDSS DR4.  With the respect to the listed
conditions we have selected 321516 objects totally.

The angular coverage of the survey is not uniform but observations
have been done in different sky regions. For this reason we have
considered three rectangular angular fields (named R1, R2 and R3) in
the SDSS internal angular coordinates $(\eta,\lambda)$: in such a way we
do not have to consider the irregular boundaries of the survey mask,
as we have cut such boundaries to avoid uneven edges of observed
regions. In Tab.\ref{tbl_VLSamplesProperties2} we report the
parameters of the three angular regions we have considered. In
addition we do not use corrections neither for the redshift
completeness mask nor for the fiber collision effects. Completeness
varies mostly nearby the current survey edges which are excluded in our
samples. Fiber collisions in general do not represent a problem for
measurements of galaxy correlations (see discussion in, e.g., Strauss
et al., 2002).

\begin{table}
\begin{center}
\begin{tabular}{|c|c|c|c|c|c|}
  \hline
  Region name & $\eta_1$ & $\eta_2$ & $\lambda_1$ & $\lambda_2$ & $\Omega$ \\
  \hline
    R1    & 9.0    & 36.0  & -47.0&  8.0 & 0.41 \\
    R2    & -33.5  & -16.5 &-54.0 & -24.0& 0.12   \\
    R3    & -36.0  & -26.5 & 2.5  &  43.0& 0.11  \\
  \hline
\end{tabular}
\end{center}
\caption{Main properties of the angular regions considered: 
The limits in degrees of the cuts are chosen using the intrinsic
coordinates of the survey $\eta$ and $\lambda$ (in degrees). The last column
$\Omega$ gives the solid angle of three angular regions in
steradians.}
\label{tbl_VLSamplesProperties2}
\end{table}

\subsection{Construction of VL samples}

To construct VL samples which are unbiased for the selection effect
related to the cut in the apparent magnitude, we have applied a
standard procedure (see e.g.  Zehavi et al., 2004): First of all we
compute metric distances as
\begin{equation}
\label{MetricDistance} 
r(z) = \frac{c}{H_0}
\int_{\frac{1}{1+z}}^{1} {\frac{dy}{y \cdot
\left(\Omega_M/y+\Omega_\Lambda \cdot y^2 \right)^{1/2}}} \; ,
\end{equation}
where we have used the standard cosmological parameters 
$\Omega_M=0.3$ and $\Omega_\Lambda=0.7$ with $H_0=100 h$ km/sec/Mpc.

We use Petrosian apparent magnitudes in the $r$ filter $m_r$ which are
corrected for galactic absorption. The absolute magnitudes can be
computed as
\begin{equation}
\label{AbsoluteMagnitude}
 M_r = m_r - 5 \cdot \log_{10}\left[r(z) \cdot (1+z)\right] - K_r(z) - 25.
\end{equation}
where $K_r(z)$ is the K-correction. As the redshift range considered
is small from a cosmological point of view (i.e. $z \leq 0.3$), to
estimate the K-corrections $K_r(z)$ (linearly proportional to $z$ and
thus small in this context) we have used the simple interpolating
formula
\begin{equation}
\label{K-CorrDef}
  K_r(z) = (2.61 \cdot (m_g-m_r)-0.64) \cdot z \;,
\end{equation} 
where $m_g$ is the apparent magnitude in the $g$ filter. This
corresponds to the calculated K-corrections in Blanton et al. (2001
--- see their Fig.4). By knowing the intrinsic $g-r$ color and the
redshift one may directly estimate the K-correction term.

We have considered 4 different VL samples (named VL1, VL2, VL3 and
VL4) defined by two chosen limits in absolute magnitude and metric
distance, whose parameters are reported in
Tab.\ref{tbl_VLSamplesProperties1}. While VL1 and VL2 actually contain
relatively faint galaxies in the local universe, VL3 sample covers a
wide range of distances, and VL4 consists of bright galaxies at
distances up to 600 Mpc/h. Considering the three different rectangular
areas (described above), in summary we have $4 \times 3 = 12$ VL
subsamples whose characteristics are reported in
Tab.\ref{tbl_VLSamplesProperties3}.  The comparison between VL samples
with the same magnitude and distance cuts, in different sky regions,
will allow us to test the statistical stationarity of galaxy distributions 
in these samples and to estimate sample-to-sample fluctuations. 

\begin{table}
\begin{center}
\begin{tabular}{|c|c|c|c|c|c|}
  \hline
  VL sample & $r_{min}$ & $r_{max}$ & $M_{min}$ 
& $M_{max}$ & $\langle \Lambda \rangle$\\
  \hline
    VL1    & 50  & 135 & -19.0 & -18.0  & 1.7  \\
    VL2    & 50  & 200 & -21.0 & -19.0  & 1.3 \\
    VL3    & 100 & 500 & -23.0 & -21.0  & 2.9  \\
    VL4    & 150 & 600 & -23.0 & -22.0  & 6  \\
   \hline
\end{tabular}
\end{center}
\caption{Main properties of the obtained VL samples: 
$r_{min}$, $r_{max}$ (in Mpc/h) are the chosen limits for the metric
distance; ${M_{min}, \,M_{max}}$ define the interval for the absolute
magnitude in each sample. The quantity $\langle \Lambda \rangle$ (in
Mpc/h) is the average distance between nearest-neighbor galaxies.}
\label{tbl_VLSamplesProperties1}
\end{table}
\begin{table}
\begin{center}
\begin{tabular}{|l|c|c|}
  \hline
  VL Sample & N & $R_c$ \\
  \hline
  R1VL1 & 3130 &15	\\
  R1VL2 & 15181&21	\\
  R1VL3 & 27975&54	\\
  R1VL4 & 6742 &65	\\
  R2VL1 & 790  &10	\\
  R2VL2 & 3912 &15	\\
  R2VL3 & 8586 &38	\\
  R2VL4 & 1923 &42	\\
  R3VL1 & 790  &9	\\
  R3VL2 & 2895 &12	\\
  R3VL3 & 7584 &30	\\
  R3VL4 & 1503 &36 	\\
  \hline
\end{tabular}
\end{center}
\caption{Number of galaxies in each of the VL sample.
Names are given according to the discussion in the text. The scale
$R_c$ (in Mpc/h) is discussed in Sect.\ref{rc} below.}
\label{tbl_VLSamplesProperties3}
\end{table}
%


\subsection{Nearest neighbor distribution}

The NN distance probability distribution depends on the cut
in absolute magnitude of a given VL sample. We expect this function
not to be dependent on the angular sky cuts if the distribution is
statistically stationary in the different VL samples. As discussed in
Vasilyev, Baryshev \& Sylos Labini (2006) space correlations introduce
a deviation from the case of a pure Poisson distribution: particularly
the average distance $\langle
\Lambda \rangle$ between NN is expected to be smaller than for the
Poisson case in the same sample and with the same number of
points. The measurements in the data, obtained by simple
pair-counting, are shown in Figs.\ref{FIGnn1}-\ref{FIGnn4}. When a VL
sample includes fainter galaxies (e.g. VL1,VL2) $\langle \Lambda
\rangle$ is smaller  
(see Tab.\ref{tbl_VLSamplesProperties1}) than for the case when only
brighter galaxies are inside (e.g. VL3,VL4). This is because brighter
galaxies are sparser than fainter ones. This corresponds to the
exponential decay of the galaxy luminosity function at the bright end
(see discussion in Gabrielli et al., 2004)
\begin{figure}
\begin{center}
\includegraphics*[angle=0, width=0.5\textwidth]{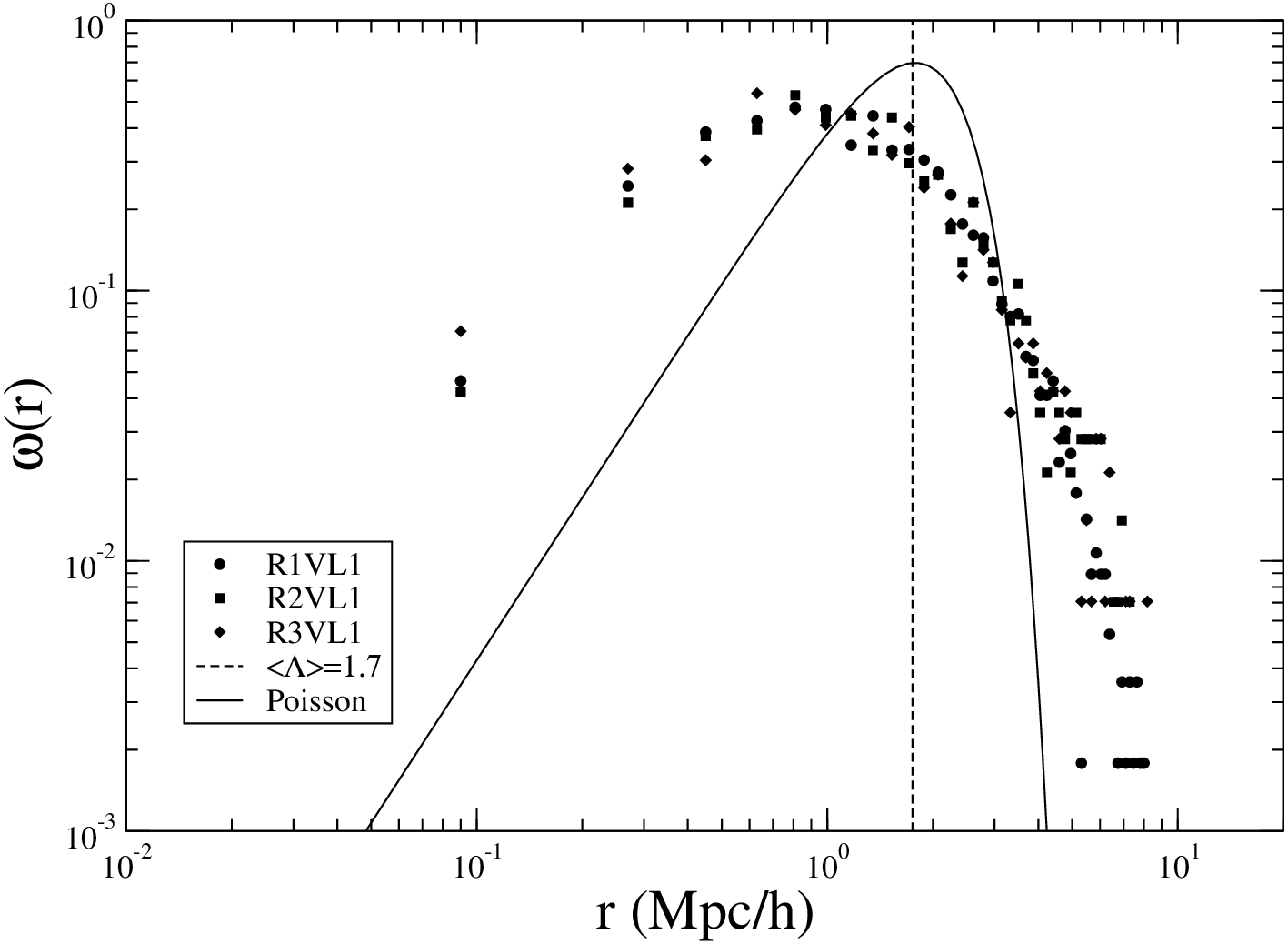}
\end{center}
\caption{Nearest Neighbor distribution in  VL1 sample: different symbols 
correspond to different angular regions. The average distance between 
nearest galaxies is $\langle \Lambda \rangle = 1.7$  Mpc/h. 
For reference the solid line represents the NN distribution for a 
Poisson configuration with the {\it same}  $\langle \Lambda \rangle$:
one may notice that the tails of this function decay more rapidly.} 
\label{FIGnn1}
\end{figure}
\begin{figure}
\begin{center}
\includegraphics*[angle=0, width=0.5\textwidth]{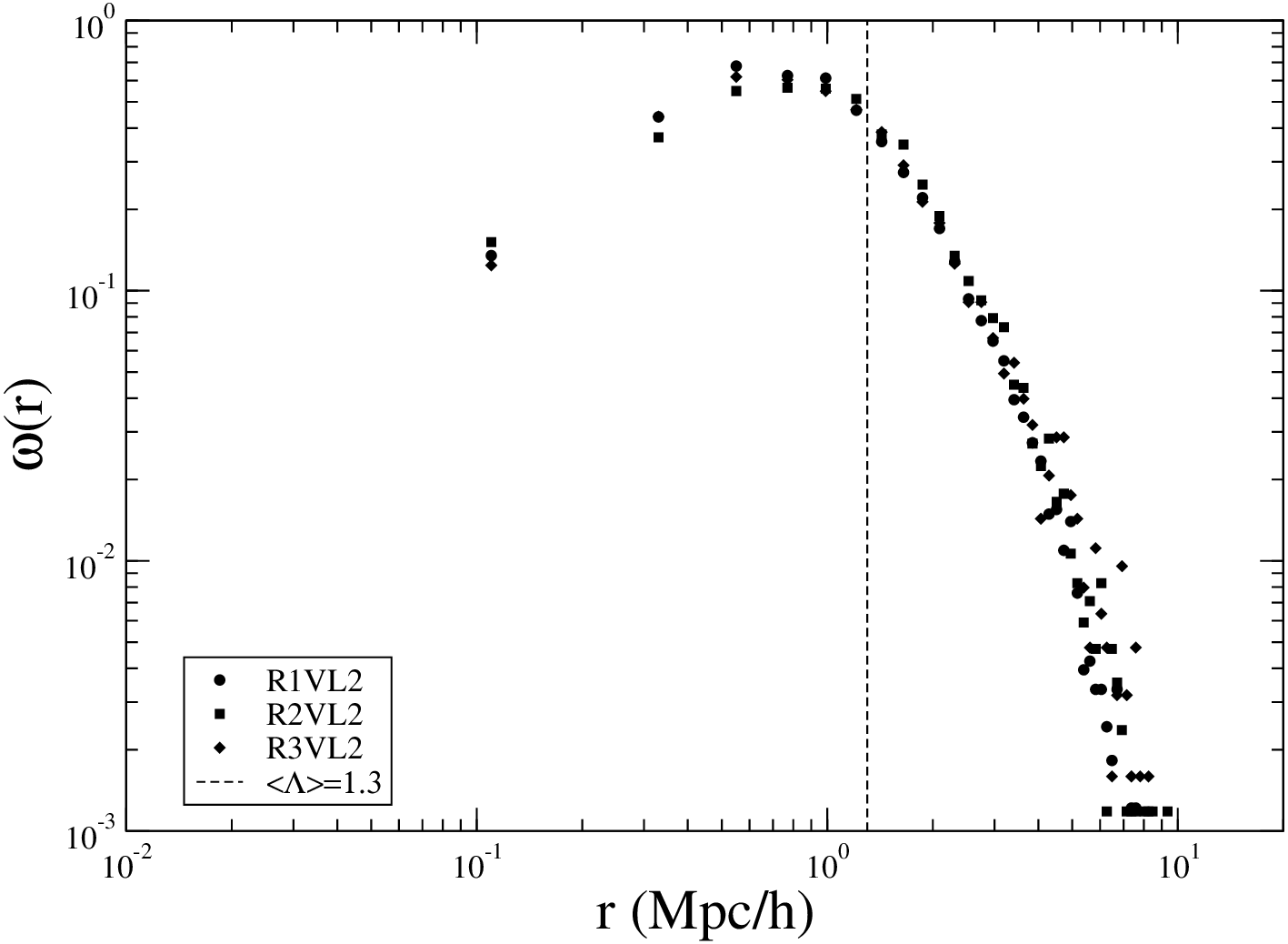}
\end{center}
\caption{As Fig.\ref{FIGnn1} but for the 
VL2 samples. The average distance between galaxies is $\langle \Lambda
\rangle = 1.3$ Mpc/h. 
}
\label{FIGnn2}
\end{figure}
\begin{figure}
\begin{center}
\includegraphics*[angle=0, width=0.5\textwidth]{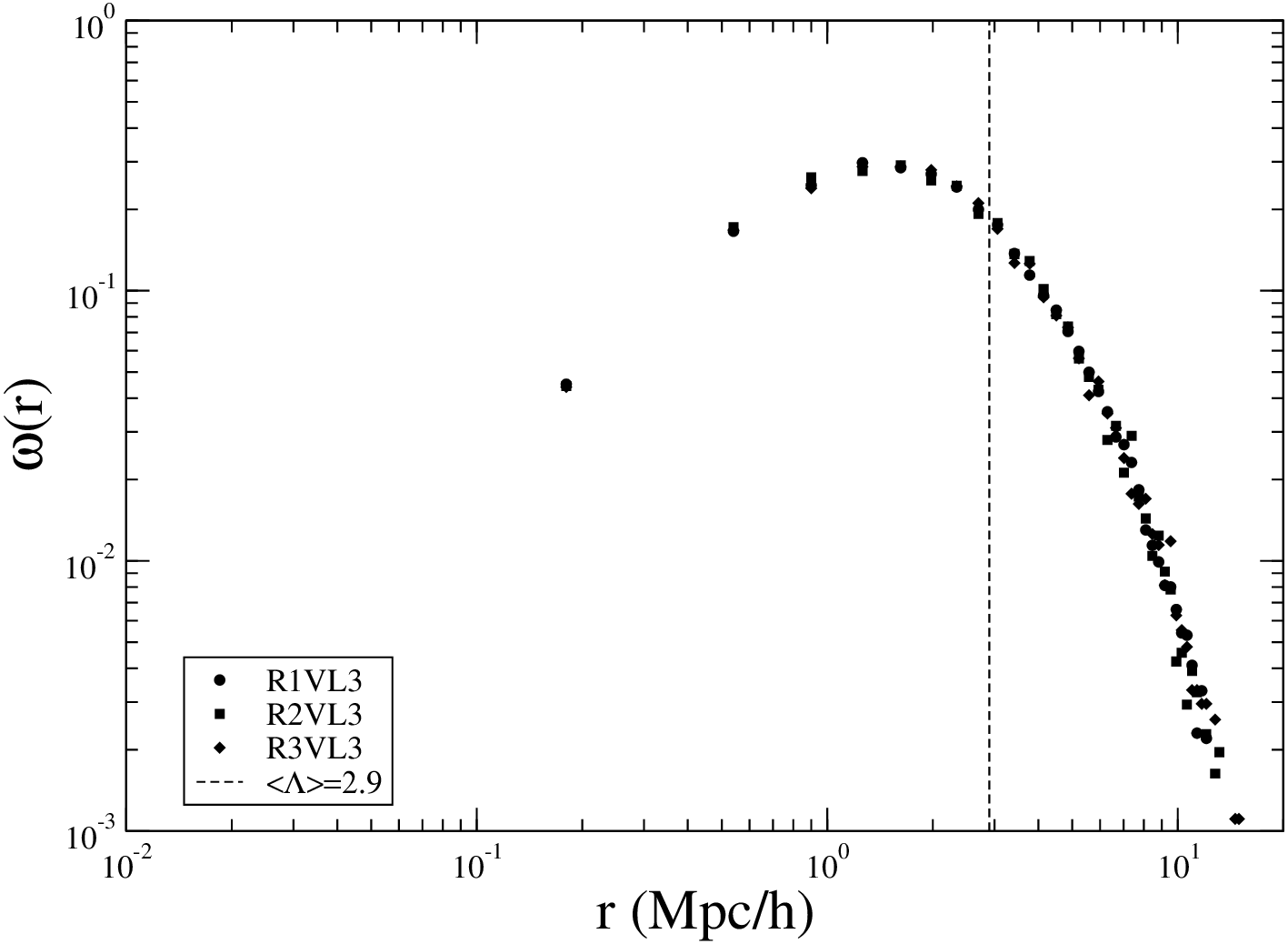}
\end{center}
\caption{As Fig.\ref{FIGnn1} but for the 
VL3 samples.  The average distance between nearest galaxies is
$\langle \Lambda \rangle = 2.9$ Mpc/h.}
\label{FIGnn3}
\end{figure}
\begin{figure}
\begin{center}
\includegraphics*[angle=0, width=0.5\textwidth]{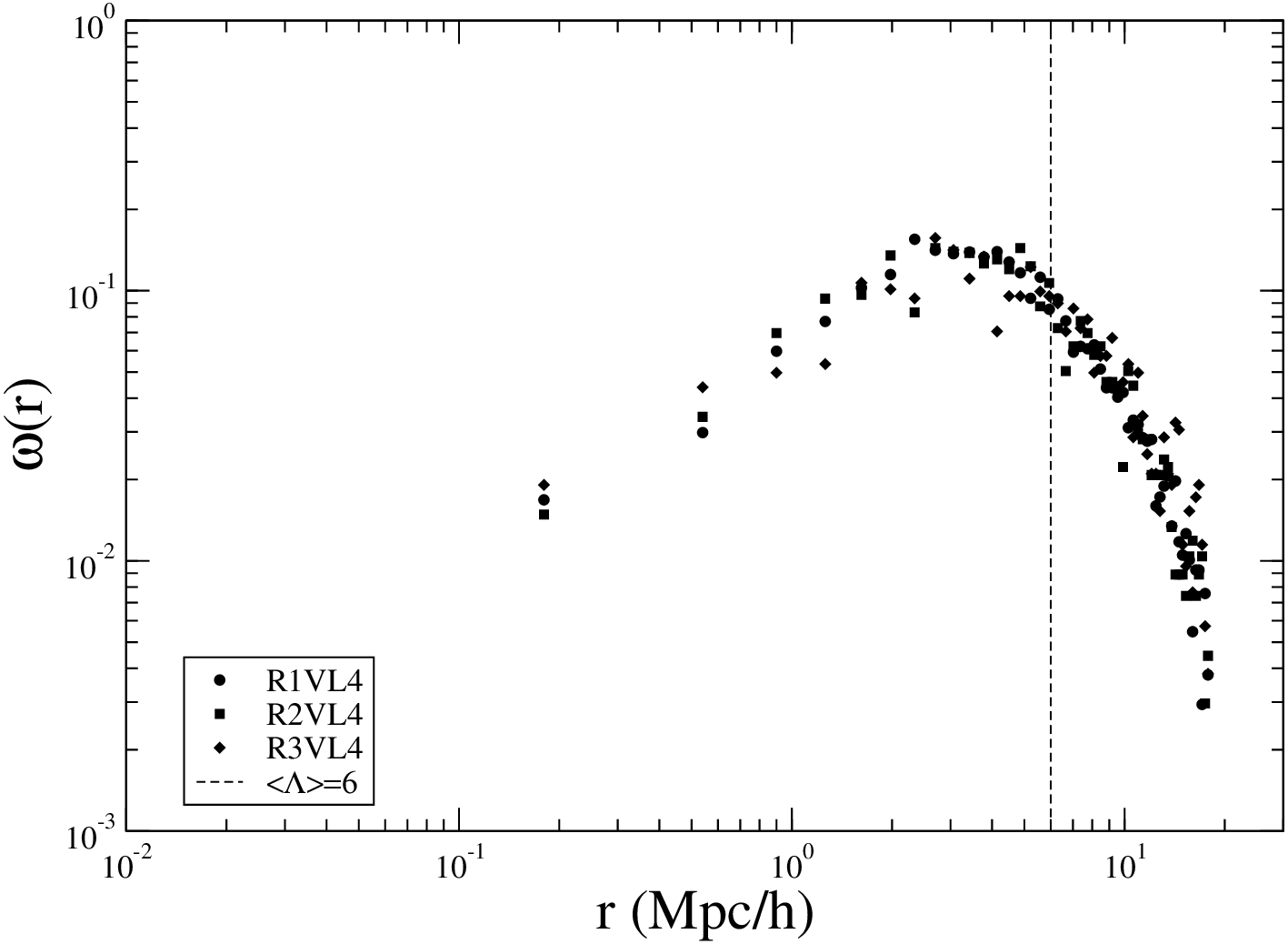}
\end{center}
\caption{
As Fig.\ref{FIGnn1} but for the VL4 samples.The average distance
between nearest galaxies is $\langle \Lambda \rangle = 6$ Mpc/h.
}
\label{FIGnn4}
\end{figure}

Note that Zehavi et al. (2004) have estimated that at scale of order
$1 \div 2$ Mpc/h there is a departure from a power law behavior in the
reduced correlation function.  At the light of the discussion above we
stress that this change occurs in a range of scales where NN
correlations are dominant in all samples considered.  For the
interpretation of this behavior one may consider the relation between
the conditional density, or the reduced correlation function, and the
NN probability distribution (see Baertschiger \& Sylos Labini 2004 for
a discussion of this point). In this respect, in the comparison of
galaxy data with N-body simulations, one has to be careful in that
these small-scale properties can be determined by sampling, sparseness
and other more subtle finite size effects related to the precision of
a given N-body simulation (Baertschiger \& Sylos Labini 2004).

We have then studied the effect of the fiber collisions on the NN
statistic: about $6\%$ of galaxies that satisfy the selection criteria
of the main galaxy sample are not observed because they have a
companion closer than the 55 arc-sec minimum separation of
spectroscopic fibers (Strauss et al., 2002). However not all
55-arc-sec pairs are affected by fiber collisions, because some of the
SDSS were observed spectroscopically more than once. We have
identified all $<=55$ arc-sec pairs for which both galaxies have
redshifts, and we have randomly removed one of those redshifts in each
case to make a new sample with an even more severe fiber collision
problem than the existing sample. Because of the very small number of
galaxy pairs with angular separation $<= 55$ arc-sec (of order of few
percent in all the volume limited samples we have considered) there is
no sensible effect of the results. In fact, for galaxies in the main
sample the average redshift $z \sim 0.1$, and hence the angular
distance 55 arc-sec corresponds to the linear separation $r \sim 0.1$
Mpc/h which is marginally outside the scales interval we have studied
the NN distribution, i.e. $r>0.2$ Mpc/h.  Hence we expect that the
fiber collision effect does not influence our results as indeed we
find.


\subsection{Number counts in VL samples} 

A simple statistics which can be easily computed in VL samples is
represented by the differential number counts. This gives us a first
indication about (i) the slope of the counts and (ii) the nature of
fluctuations (see e.g. Gabrielli et al.  2004). In general we
may write that the number of points counted from a given point chosen
as origin (in this case the Earth) grows as
\be
N(r) \sim r^D \;.
\ee
This represents the radial counts in a spherical volume of radius $r$
around the observer (or in a portion of a sphere). In the case $D=3$ the
distribution is uniform and $D<3$ if it is, for example, fractal or if
there is a systematic effect of depletion of points as a function of
distance. In this situation we neglect relativistic effects, which are
anyway small in the range of redshift considered. However as noticed by
Gabrielli et al. (2004) these corrections may change the
slope of the counts but not the intrinsic fluctuations.  

 Given that a VL sample is defined by two cuts in distance we compute
\be
\label{e2} 
n(r) = \frac{d N(r)}{dr} \sim r^{D-1} \;,
\ee
i.e. the differential number counts in shells. Simply stated we expect
the exponent in Eq.\ref{e2} to be 2 when the distribution is uniform;
in this case we also expect to see small (normalized) fluctuations
generally decaying as the volume or faster for the case of
super-homogeneous case (i.e. for standard cosmological density fields
--- see discussion in  Gabrielli et al., 2004)

Results in the samples considered are shown in
Figs.\ref{FIGnc1}-\ref{FIGnc4}, where for each sample we have
normalized the counts to the solid angle of the corresponding angular
region. One may note that the best fit exponent (reported in the
figures) fluctuates, and in several case it is larger than 2. This
means that there are large fluctuations as evidenced by the non-smooth
behaviors of $n(r)$ in the different samples. A similar evidence of
the effect of large scale structures in these samples on other
statistical quantities has been recently pointed out by Nichol et
al. (2006).

This is a first rough indication that the question of uniformity at
scales of order 100 Mpc/h is not simple to be sorted out in these
samples. These large fluctuations in slope and amplitude correspond to
the presence of large scale galaxy structures extending up to the
boundaries of the various samples considered. We do not present a more
quantitative discussion of these behaviors as the statistics is rather
weak. 
\begin{figure}
\begin{center}
\includegraphics*[angle=0, width=0.5\textwidth]{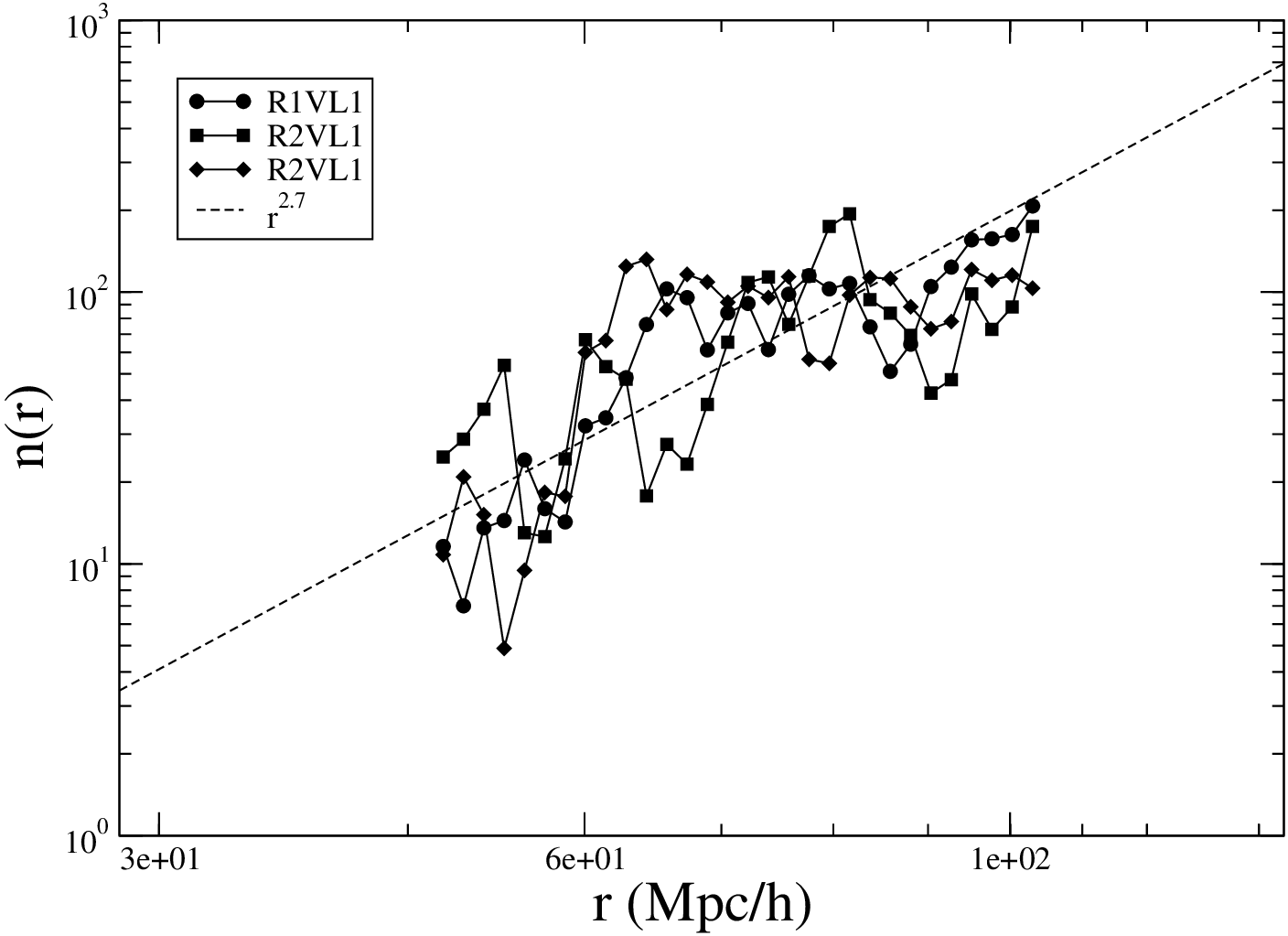}
\end{center}
\caption{Differential number counts as a function of distance
in the VL1 sample in different angular regions normalized to their own
solid angle}
\label{FIGnc1}
\end{figure}
\begin{figure}
\begin{center}
\includegraphics*[angle=0, width=0.5\textwidth]{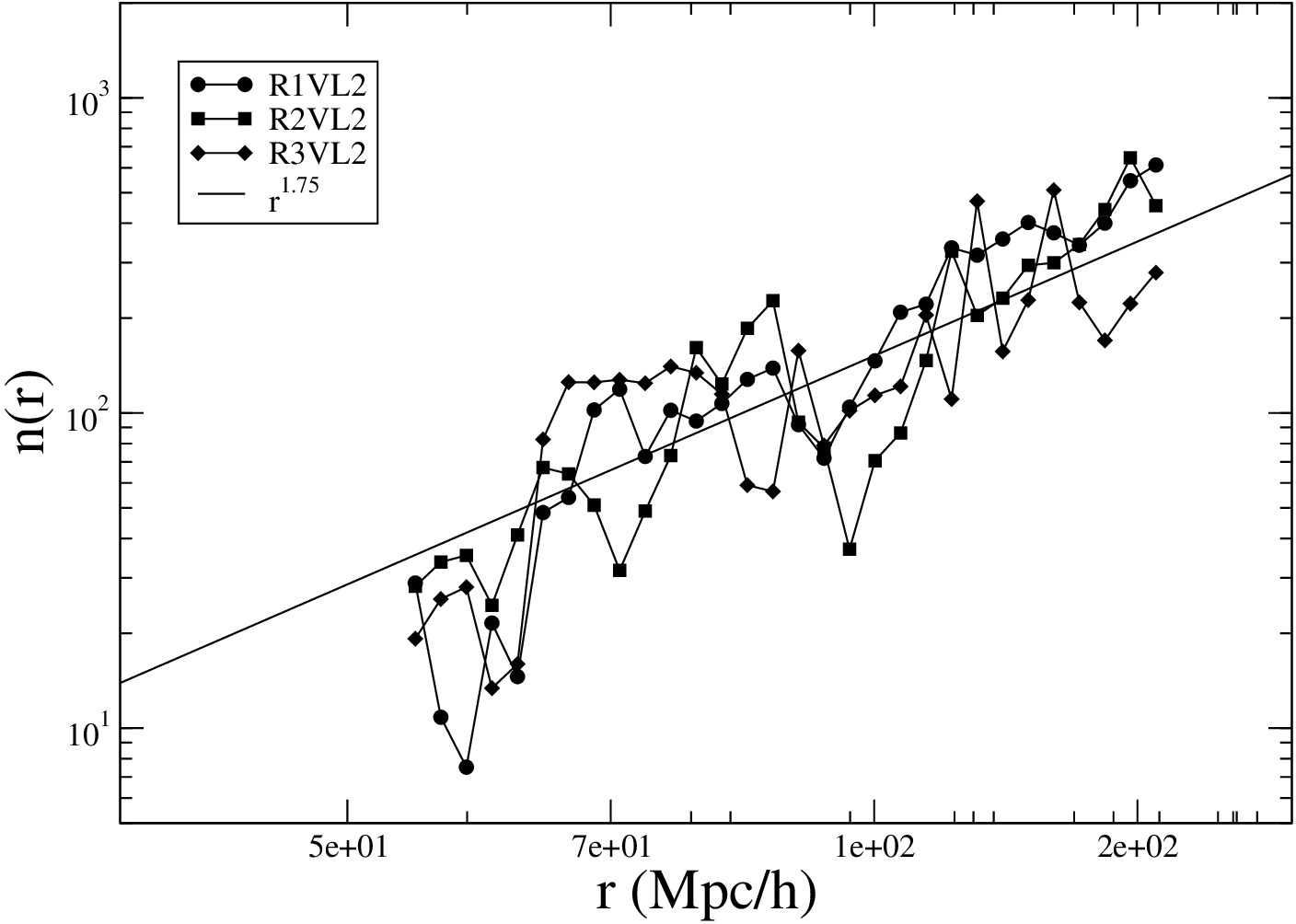}
\end{center}
\caption{
The same as Fig.\ref{FIGnc1} but for the 
VL2 samples}
\label{FIGnc2}
\end{figure}
\begin{figure}
\begin{center}
\includegraphics*[angle=0, width=0.5\textwidth]{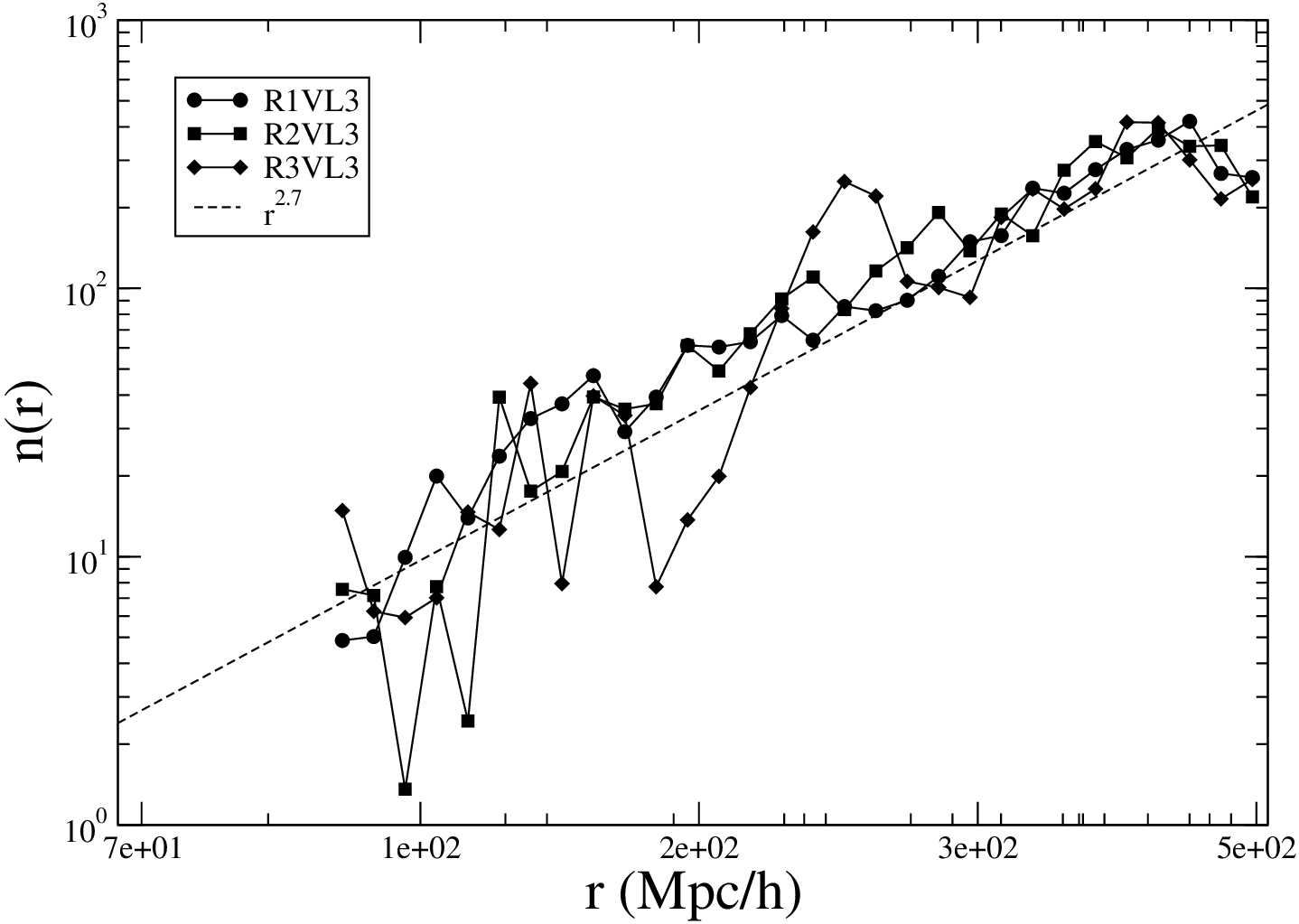}
\end{center}
\caption{The same as Fig.\ref{FIGnc1} but for the 
VL3 samples}
\label{FIGnc3}
\end{figure}
\begin{figure}
\begin{center}
\includegraphics*[angle=0, width=0.5\textwidth]{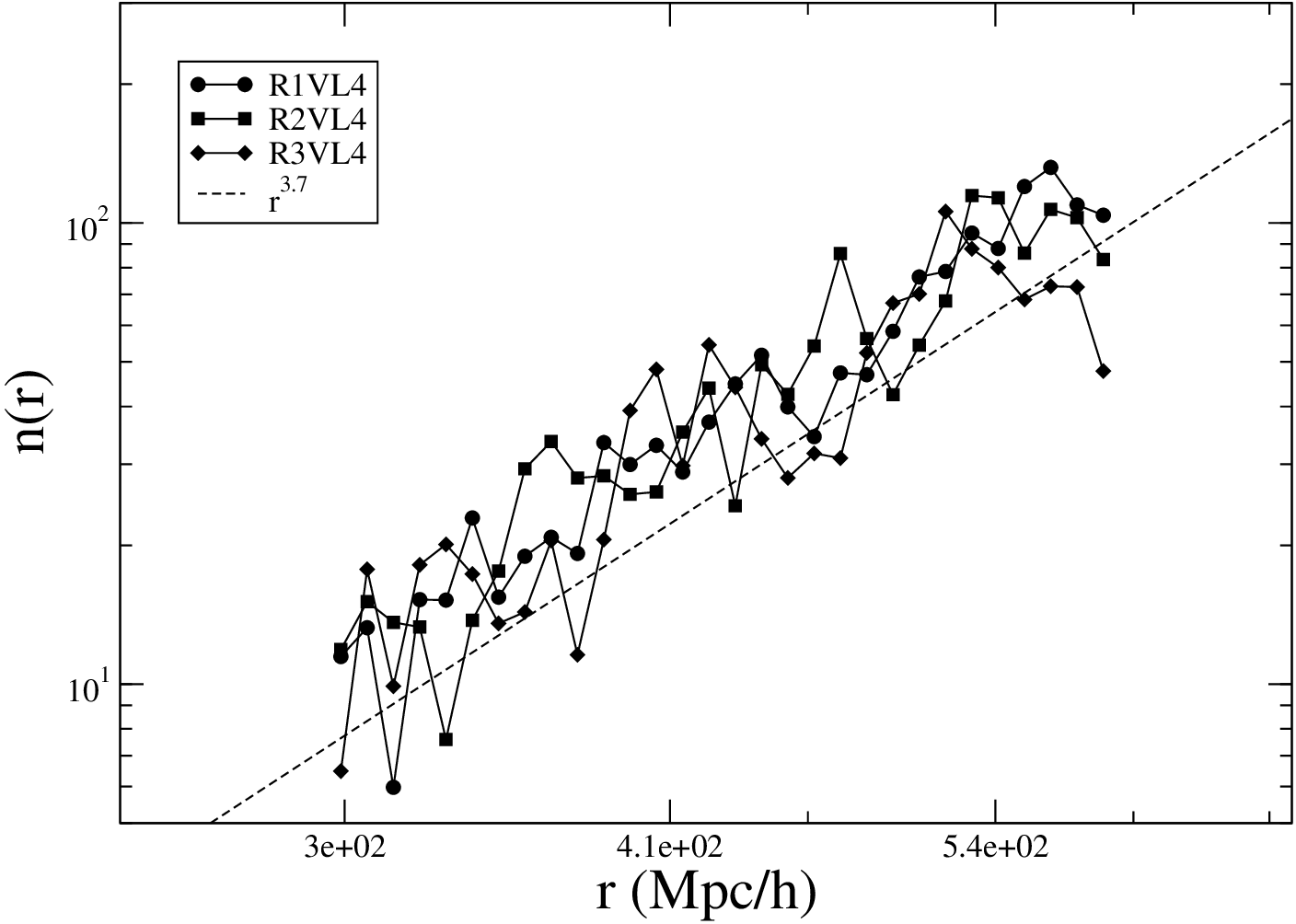}
\end{center}
\caption{The same as Fig.\ref{FIGnc1} but for the 
VL4 samples}
\label{FIGnc4}
\end{figure}


\section{Correlation properties of galaxy distributions} 

We study now the behavior of the conditional average density in the
various VL samples discussed in the previous section. We use the
full-shell estimator, discussed extensively in Gabrielli et al. (2004)
and recently in Vasilyev, Baryshev \& Sylos Labini (2006). This
estimator has the advantage of making no assumption in the treatment
of boundary conditions and it is the more conservative one among
estimators of two-pint correlations (see discussion in
Kerscher,1999). Briefly the conditional density in spheres
$\langle n(r)^*\rangle_p$ is defined for an ensemble of realizations
of a given point process, as
\begin{equation}\label{Gamma*-r}
 \langle n(r)^*\rangle_p
 = \frac{\langle{N(r)}\rangle_{p}}{V(r)}. 
\end{equation}
This quantity measures the average number of points
$\langle{N(r)}\rangle_{p}$ contained in a sphere of volume
$V(r)=\frac{4}{3}\pi{r}^{3}$ with the condition that the center of the
sphere lies on an occupied point of the distribution (and
$\langle{...}\rangle_{p}$ denotes the conditional ensemble average).
Such a quantity can be estimated\footnote{For simplicity we use the
same symbol for the ensemble average and for the estimator of all
statistical quantities defined in this section} in a finite sample by
a volume average (supposing ergodicity of the point distribution)
\begin{equation}
\label{Gamma*E-r}
 \langle n(r)^*\rangle_p
 = \frac{1}{N_c(r)} \sum_{i=1}^{N_c(r)}{\frac{N_i(r)}{V(r)}},
\end{equation}
where $N_c(r)$ --- the number of points  for which, when chosen 
as centers of a sphere of radius $r$, this is fully
contained in the sample volume --- averaging by the sample
points. (The estimation of the conditional density in shells $\langle
n(r)\rangle_p$ proceeds in the same way, except for the fact of
considering spherical shells instead of spheres centered on the points
--- see e.g. Vasilyev, Baryshev \& Sylos Labini 2006).

Therefore this full-shell estimator has an important constraint: it is
measured only in spherical volumes fully included in the sample
volume.  In this situation the number of centers $N_c(r)$ over which
the average Eq.\ref{Gamma*E-r} is performed, becomes strongly
dependent on the scale $r$ when $r \rightarrow R_s$, being $R_s$ the
sample size. In this context such a length scale can be defined as the
radius of the largest sphere fully included in the sample volume: the
center of such a sphere lies in the middle of the sample volume.

Thus, when approaching the scale $R_s$ there are two sources of
fluctuations which increase the variance of the measurements.  From the
one hand the number of points over which the average is performed
decreases very rapidly and from the other hand the remaining points
are concentrated toward the center of the sample. In such a way
systematic fluctuations may affect the estimation, given that these
are not averaged out by the volume average. An estimation of the scale
beyond which systematic effects become strong  and thus important.

The following subsection is focused  to the discussion of the
measurements of $\langle n(r)^* \rangle_p$ in the different VL samples,
while Sect.\ref{rc} is devoted to the problem of the determination of
the maximum scale up to which the volume average is properly
performed, and thus beyond which systematic unaveraged fluctuations
may affect the behavior of the conditional density.


\subsection{Estimation of the conditional density}
\label{egamma}

The results of the measurements in redshift space of the conditional
density by the full-shell estimator, in VL samples with the same cuts
in absolute magnitude and distance but in different angular regions,
are reported in Figs.\ref{FIGgamma1}-\ref{FIGgamma4}. The formal
statistical error, reported in the figures, for the determination of
$\langle n(r)^*\rangle_p$ at each scale, can be simply derived from the
dispersion of the average
\be
\label{errgamma}
\Sigma^2(r) = \frac{1}{N_c(r)} \sum_{i=1}^{N_c(r)-1} 
\frac{\left( n(r)_i^* - \langle n(r)^*\rangle_p \right)^2}
{N_c(r)-1} \;,
\ee
where $n(r)_i^*$ represents the determination from the $i^{th}$
point. One may see that such an error is very small, except for the
last few points. However, as discussed below, when $r \rightarrow R_s$
systematic fluctuations can be more important than statistical
ones. 

One may note the following behaviors:
\begin{itemize}
\item 
In the three VL1 samples the signal is approximately the same up to 10
Mpc/h, where the conditional density has a power-law behavior
\be
\langle n(r)^*\rangle_p  \sim r^{-\gamma}
\ee
with exponent $\gamma = 1.0 \pm 0.1$. The sample R3VL1 has an $R_s$ of
order 10 Mpc/h, while the sample R1VL1 about 25 Mpc/h and R2VL1 about
15 Mpc/h. In these two former samples the signal is different in the
range of scale 10-20 Mpc/h and clearly affected by large systematic
fluctuations.

\item  For the three VL2 samples the situation 
is similar to the previous one. There is a difference in the
amplitude of R1VL2 and R2VL2 of about a factor 2. Nevertheless the
power-index is very similar in all the three samples and
$\gamma=1.0\pm0.1$. All samples present a deviation from a power-law
at their respective $R_s= 35, \; 20, \; 15$ Mpc/h. These deviations
are again a sign of finite size effects, reflecting systematic
unaveraged fluctuations, as they occur at different scales in the
three samples, but always at scales comparable to the samples size.

\item 
For the case of VL3 samples the behavior of the conditional density is
smoother at small scales: up to 30 Mpc/h all the three samples present
the same power-law correlation with an index $\gamma = 1.0 \pm
0.1$. Thus the exponent is the same as in VL1 and VL2, but, given that
$R_s$ for these samples is larger than for VL1 and VL2, it extends to
larger scales. The amplitude of the conditional density is almost the
same in the there samples up to $\sim \;$30$\div\;$40 Mpc/h. Beyond such a
scale we note that R1VL3 shows a flatten behavior, similar to the case
R2VL3 although in this case there is a deviation at large scales (from
about 40 Mpc/h). Finally the sample size for R3VL3 is about 30 Mpc/h
and thus does not give any information on the larger scales. We may
anticipate that in the following section we are going to present
several tests to clarify whether the crossover to homogeneity which
seems to be clear in the sample R1VL3 is stable in different samples
and whether systematic fluctuations are negligible.

\item The sample  VL4 is the deepest one 
 and the behavior measured is similar to VL3 although there is a clear
 difference at large scales and fluctuations are more evident. Up to
 30 Mpc/h the exponent is again $
\gamma = 1.0 \pm 0.1$, i.e. like VL1 and  VL2 at smaller scales, and VL3 at
the same scales. 
\end{itemize}
Note that the different in amplitude of the conditional density in the
different samples VL1, VL2 and VL3 is simply explained by considering
the effect of the luminosity function in the selection of the galaxies
(see Gabrielli et al., 2004 for a detailed treatment of this point).

 From this discussion we may draw our main conclusion: the correlation
 properties are independent on galaxy luminosity and they are
 characterized by a power-law index in the behavior of the conditional
 density $\gamma = 1.0 \pm 0.1$ up to 30 Mpc/h.  At larger scales, as
 shown for example in the two samples R1VL4 and R2VL4 the situation is
 less clear: fluctuations are more important because they are not
 smoothed out by the volume average.  In the next subsection we define
 the range where the volume average is properly performed.

\begin{figure}
\begin{center}
\includegraphics*[angle=0, width=0.5\textwidth]{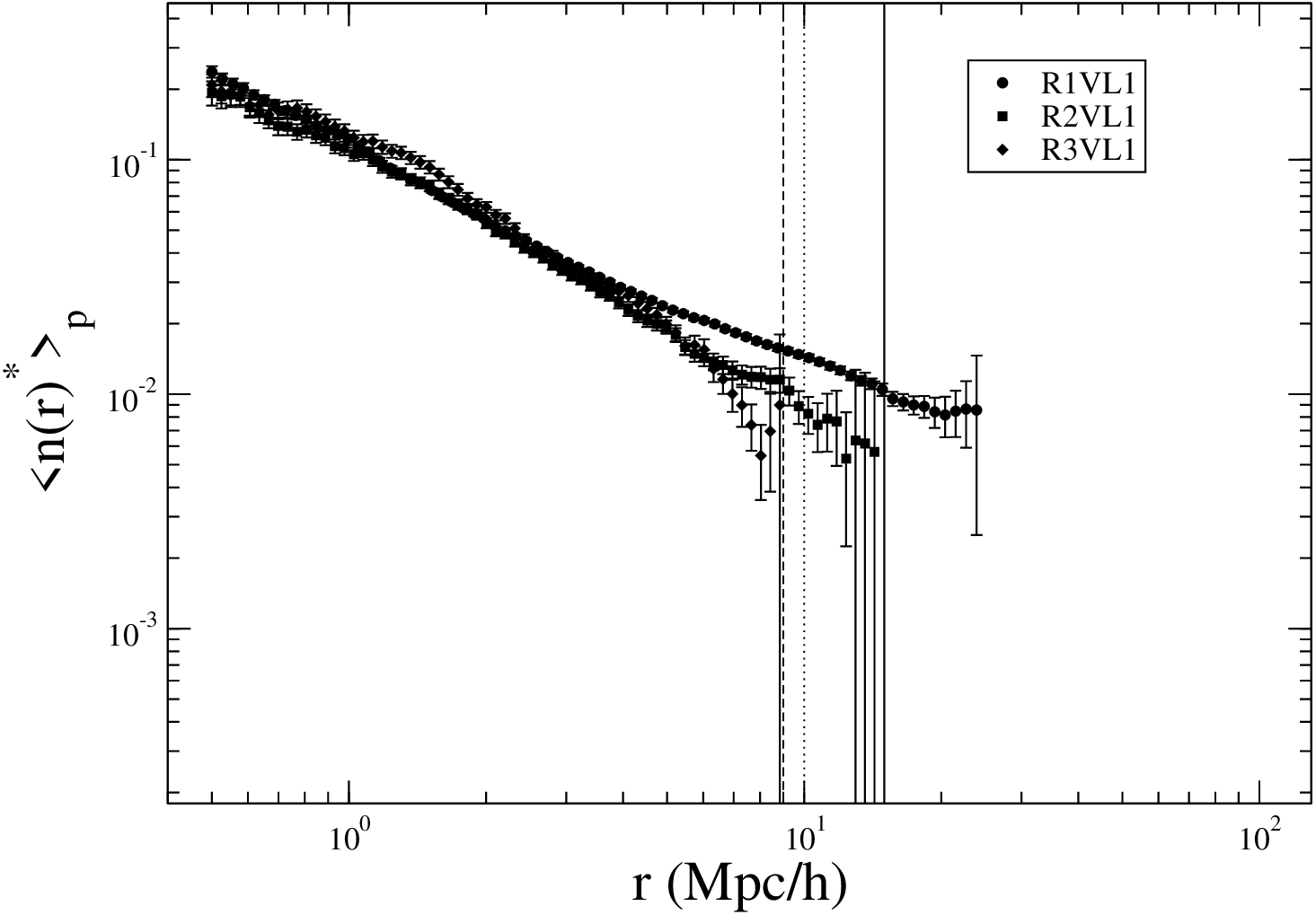}
\end{center}
\caption{Conditional density in spheres in the VL1 sample 
in the angular region R1, R2, R3. Here and in
Figs.\ref{FIGgamma2}-\ref{FIGgamma4} we report, for each sample, a
vertical line corresponding to the distance scale $R_c$ discussed in
Sect.\ref{rc} and shown in Tab.\ref{tbl_VLSamplesProperties4}
(solid-line for R1, dotted-line for R2 and dashed-line for R3)}
\label{FIGgamma1}
\end{figure}

\begin{figure}
\begin{center}
\includegraphics*[angle=0, width=0.5\textwidth]{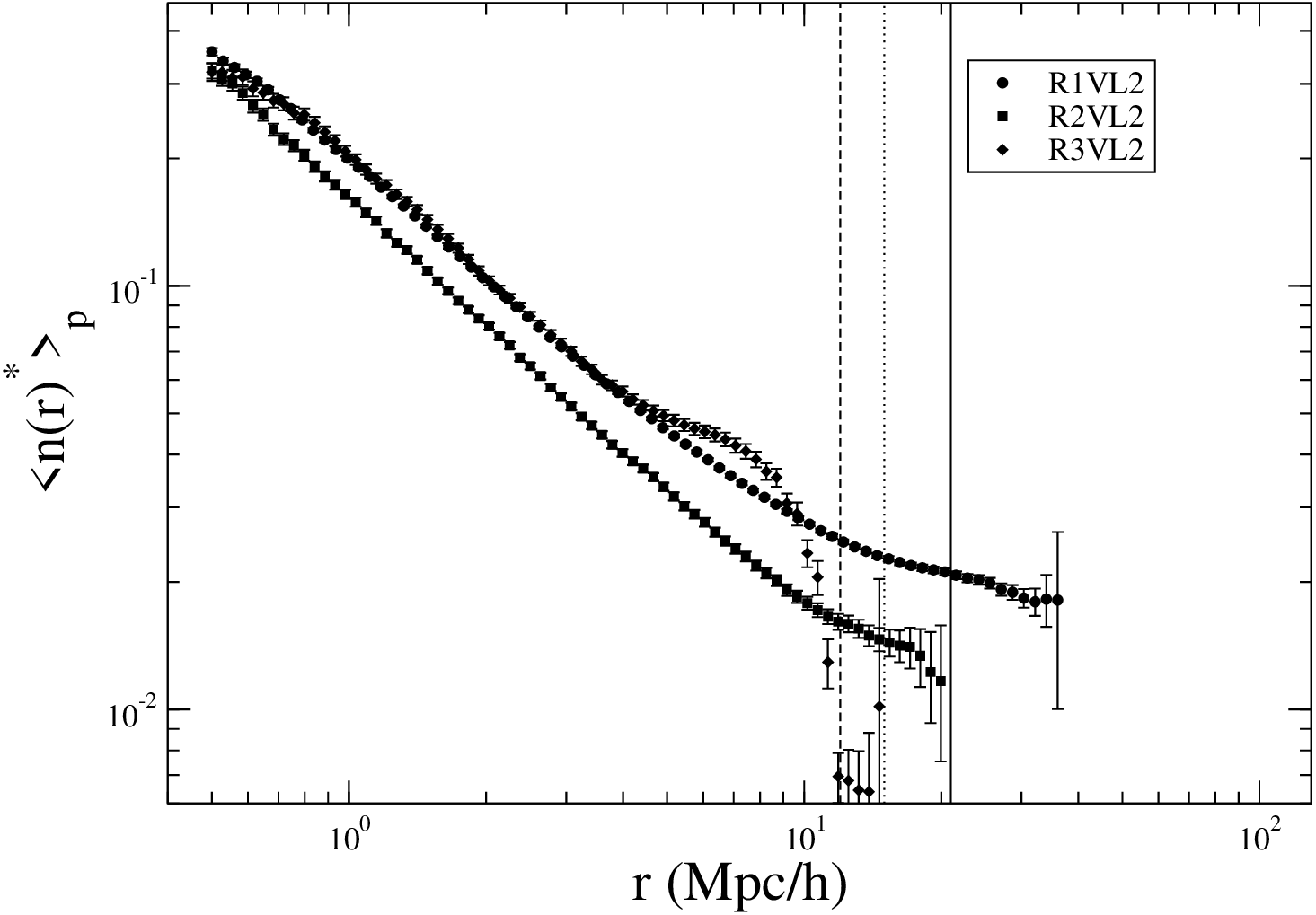}
\end{center}
\caption{As for Fig.\ref{FIGgamma1} but for  VL2 sample}
\label{FIGgamma2}
\end{figure}

\begin{figure}
\begin{center}
\includegraphics*[angle=0, width=0.5\textwidth]{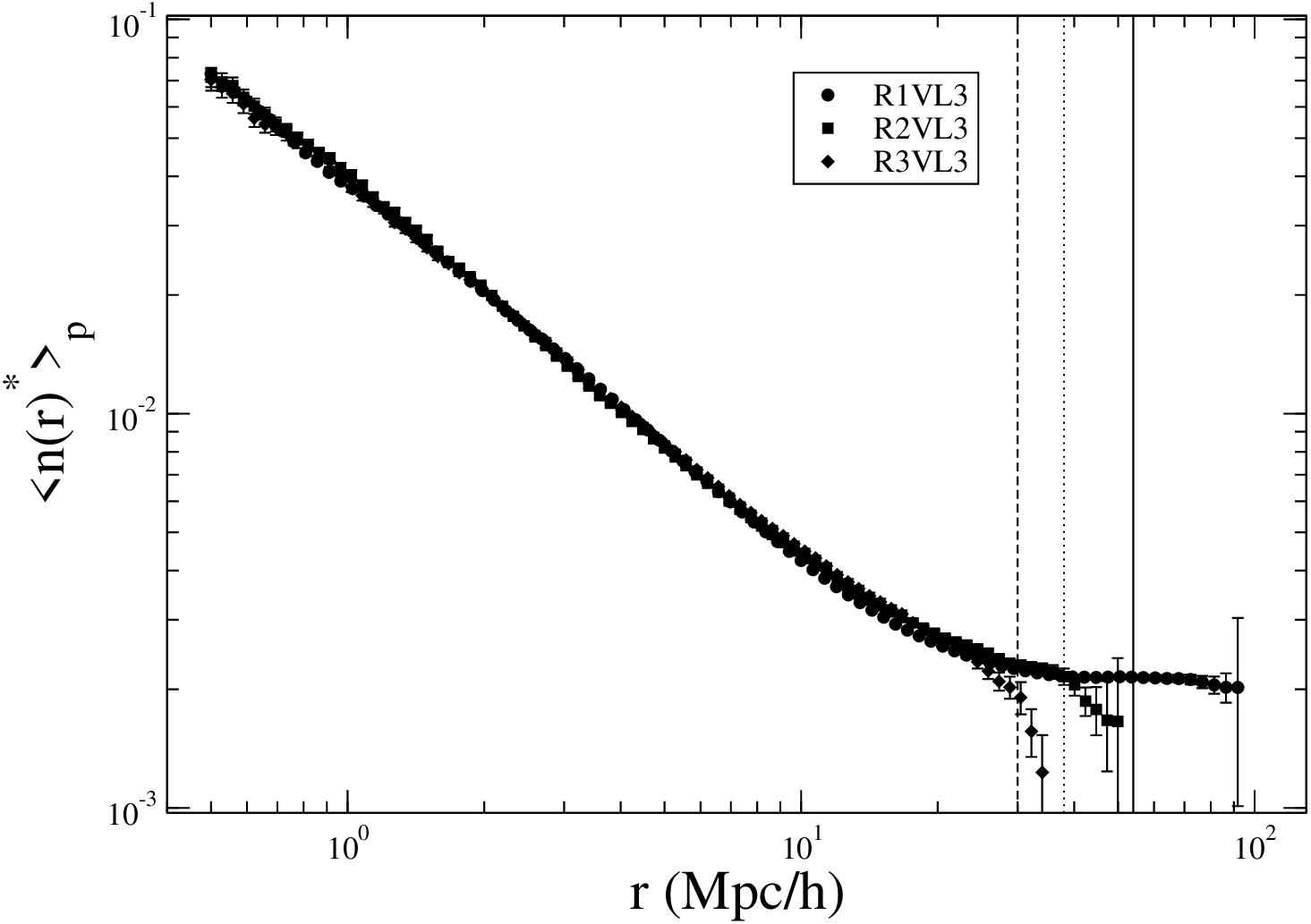}
\end{center}
\caption{As for Fig.\ref{FIGgamma1} but for  VL3 sample}
\label{FIGgamma3}
\end{figure}

\begin{figure}
\begin{center}
\includegraphics*[angle=0, width=0.5\textwidth]{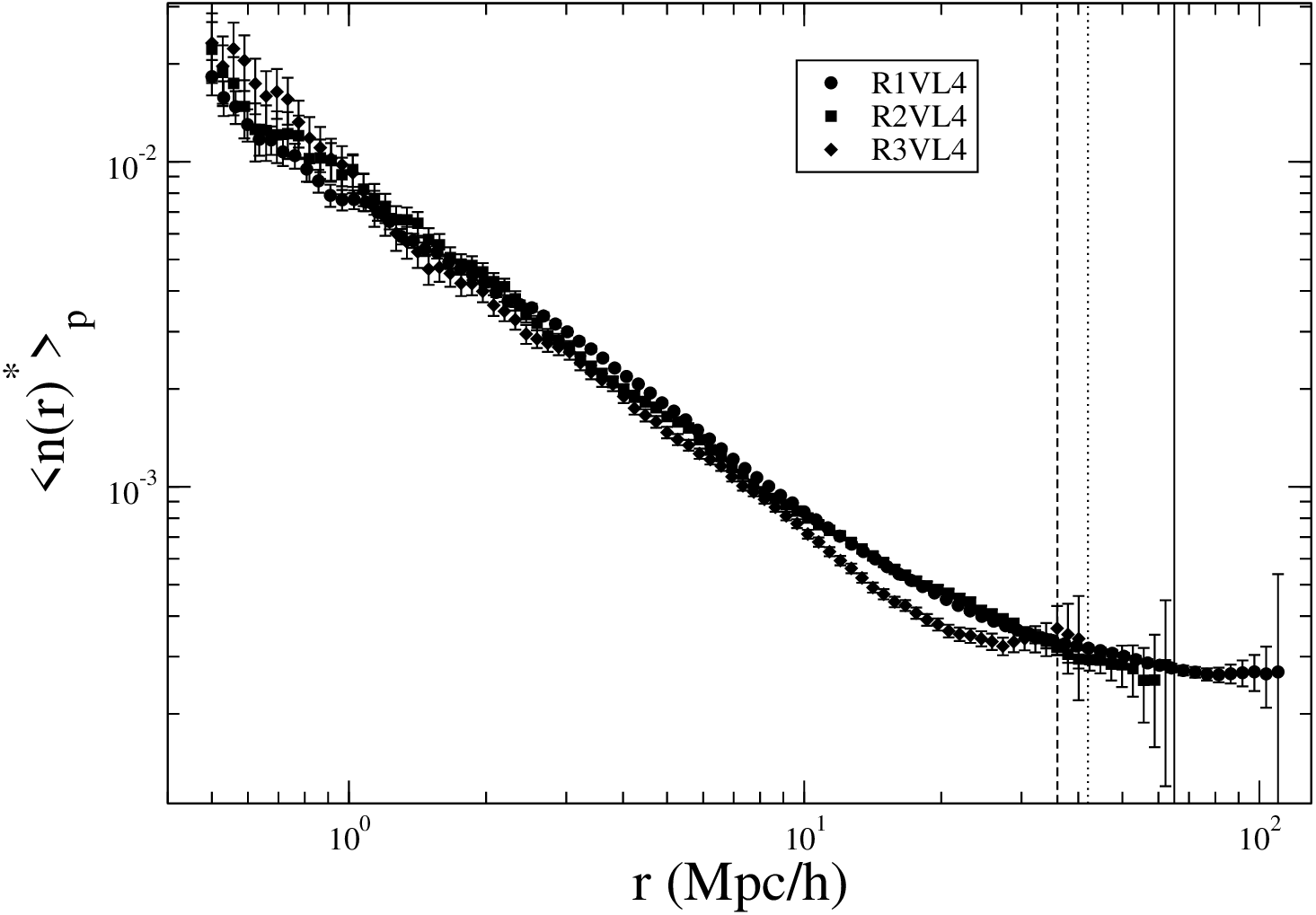}
\end{center}
\caption{As for Fig.\ref{FIGgamma1} but for  VL4 sample}
\label{FIGgamma4}
\end{figure}

%
\subsection{Finite volume effects} 
\label{rc} %

In order to quantify the finite volume effects previously mentioned,
we have divided each of the VL samples of the R1 field into two
non-overlapping contiguous angular regions, and we have recomputed the
conditional density in each of the $2 \times 4$ samples. The
properties of these subsamples are listed in
Tab.\ref{tbl_VLSamplesProperties4}. In
Figs.\ref{FIGgammaR1-1}-\ref{FIGgammaR1-4} we show the results.
\begin{table}
\begin{center}
\begin{tabular}{|c|c|c|c|c|c|}
  \hline
  Region name & $\eta_1$ & $\eta_2$ & $\lambda_1$ & $\lambda_2$ &  $N$ \\
  \hline
    R1$\_$1VL1    & 9.0    & 22.5  & -47.0 &  8.0 & 1585 \\
    R1$\_$2VL1    & 22.5   & 36.0  & -47.0 &  8.0 & 1545 \\
    R1$\_$1VL2    & 9.0    & 22.5  & -47.0 &  8.0 & 7684 \\
    R1$\_$2VL2    & 22.5   & 36.0  & -47.0 &  8.0 & 7497 \\
    R1$\_$1VL3    & 9.0    & 22.5  & -47.0 &  8.0 & 13982 \\
    R1$\_$2VL3    & 22.5   & 36.0  & -47.0 &  8.0 & 13993 \\
    R1$\_$1VL4    & 9.0    & 22.5  & -47.0 &  8.0 & 3343 \\
    R1$\_$2VL4    & 22.5   & 36.0  & -47.0 &  8.00& 3399 \\
  \hline
\end{tabular}
\end{center}
\caption{Main 
properties of the different subsamples considered in the R1 region.
The angular limits of the cuts in the intrinsic coordinates of the
survey  $\eta$ and $\lambda$ (in degrees). The last column gives the
number of points in the sample.}
\label{tbl_VLSamplesProperties4}
\end{table}
\begin{figure}
\begin{center}
\includegraphics*[angle=0, width=0.5\textwidth]{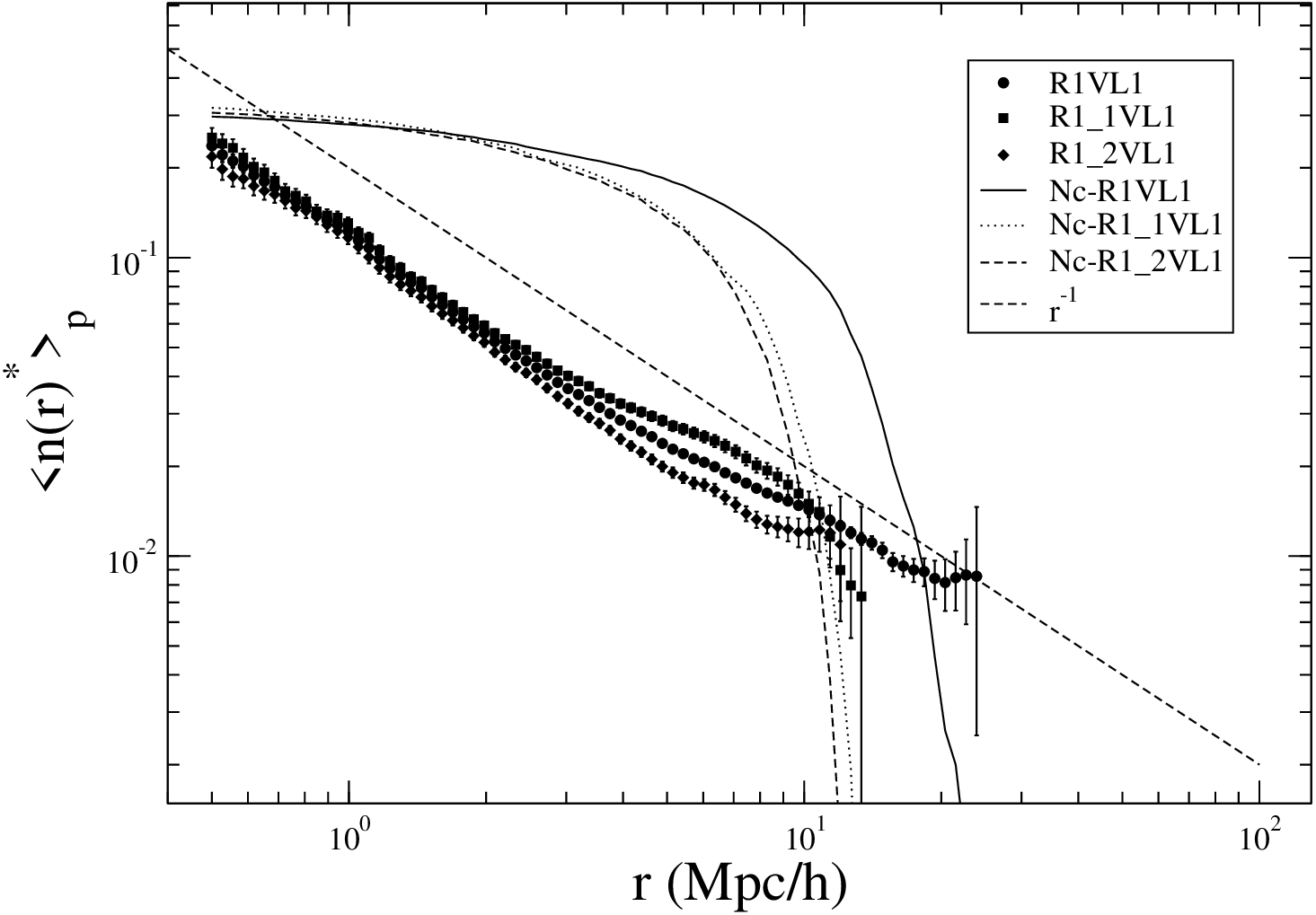}
\end{center}
\caption{Conditional density in spheres in the R1VL1 sample
and in the 2 subsamples defined by the angular cut performed as
discussed in the text. The lines labeled with $N_c$ represent the
behavior of the number of centers used in the average
(Eq.\ref{Gamma*E-r}) arbitrarily normalized. }
\label{FIGgammaR1-1}
\end{figure}
\begin{figure}
\begin{center}
\includegraphics*[angle=0, width=0.5\textwidth]{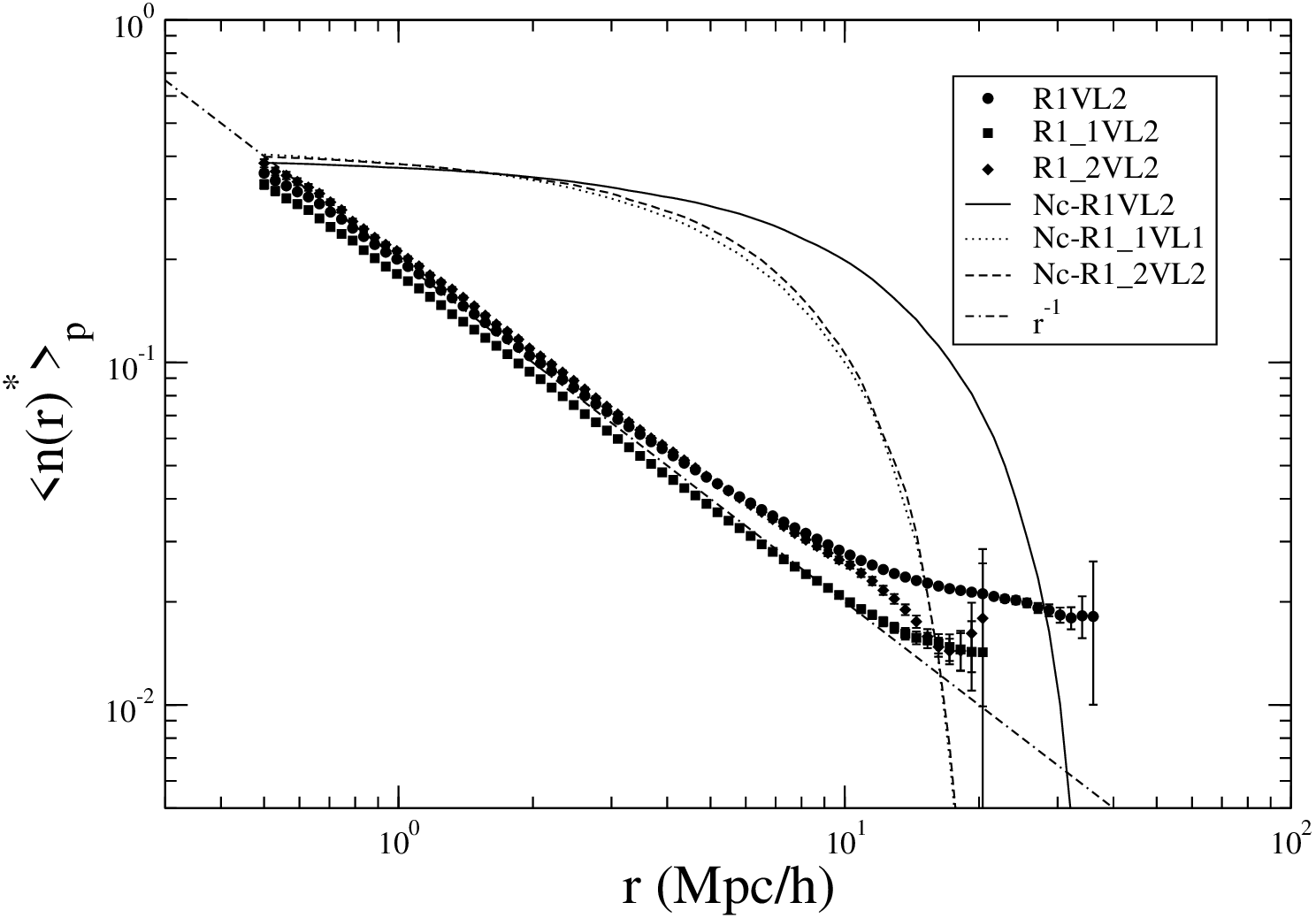}
\end{center}
\caption{As Fig.\ref{FIGgammaR1-1} but for the R1VL2 sample}
\label{FIGgammaR1-2}
\end{figure}
\begin{figure}
\begin{center}
\includegraphics*[angle=0, width=0.5\textwidth]{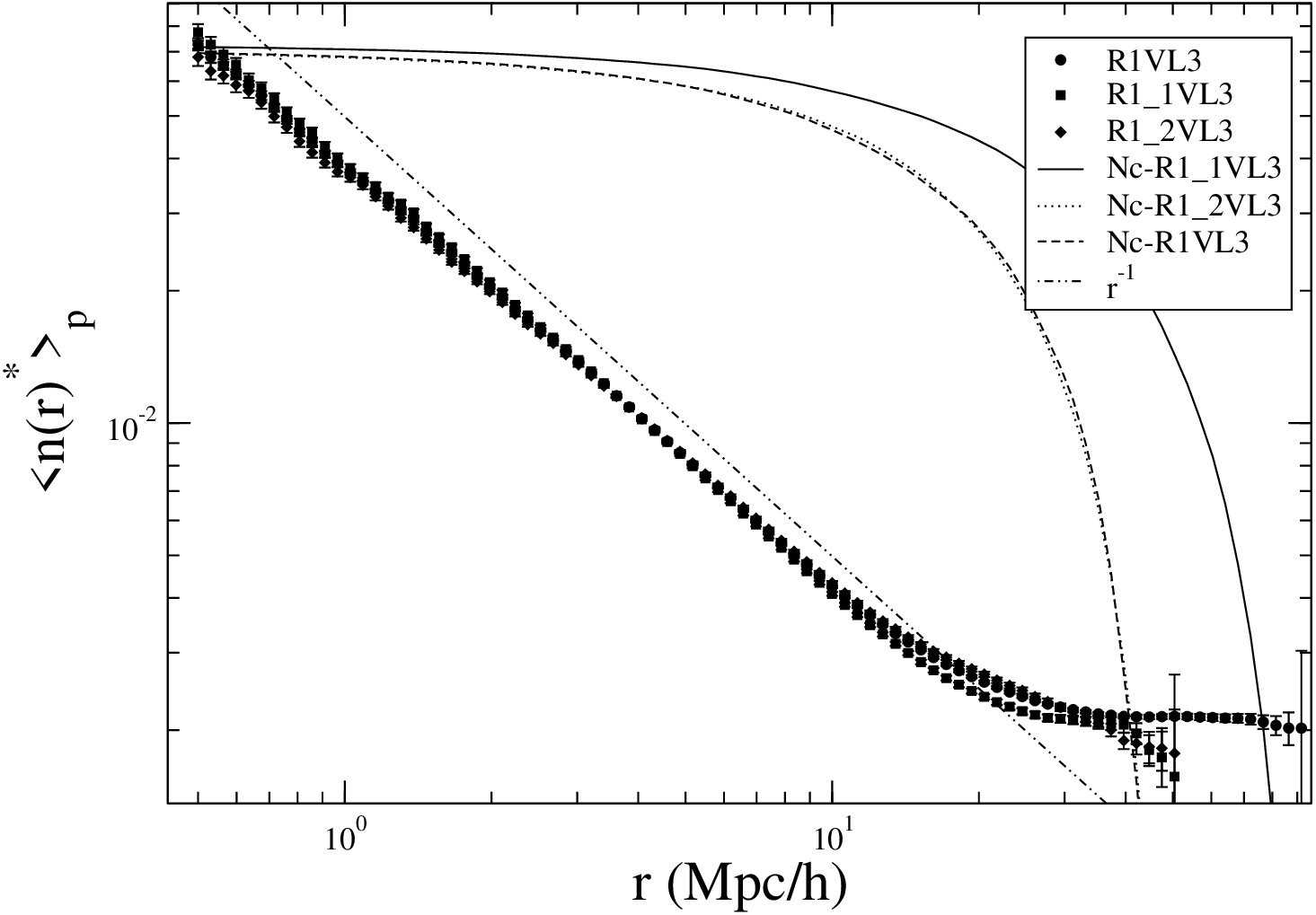}
\end{center}
\caption{As Fig.\ref{FIGgammaR1-1} but  for the R1VL3 sample}
\label{FIGgammaR1-3}
\end{figure}
\begin{figure}
\begin{center}
\includegraphics*[angle=0, width=0.5\textwidth]{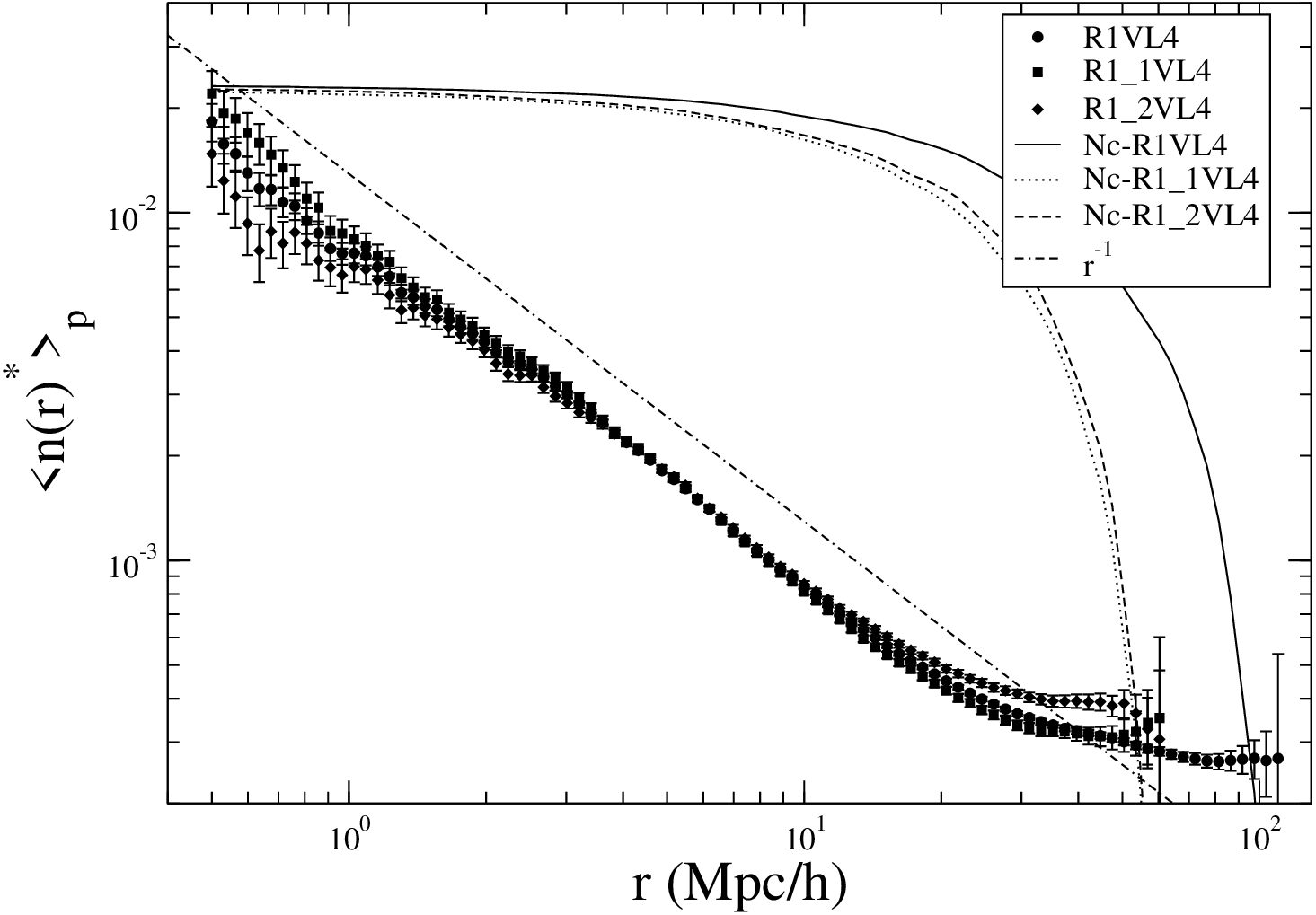}
\end{center}
\caption{As Fig.\ref{FIGgammaR1-1} but for the R1VL4 sample}
\label{FIGgammaR1-4}
\end{figure}

Let us now discuss the situation in some details.  As already mentioned
the average computed by Eq.\ref{Gamma*E-r} is made by changing, at
each scale $r$ the number $N_c(r)$ of points which do contribute. This
scale-dependency follows from the requirement that only those points
for which, when chosen as centers of a sphere of radius $r$, the
volume does not overlap or intersect the boundaries of the sample.  In
this way, in a sample of size $R_s$, when $r\ll R_s$ almost all points
will contribute to the average, while when $r\rightarrow R_s$ only
those points lying close to the center of the volume will be taken
into account in the average. Hence at large scales the average is
performed on a number of points which exponentially decays when
$r\rightarrow R_s$. In Figs.\ref{FIGgammaR1-1}-\ref{FIGgammaR1-4} we
show the behavior of the number of centers $N_c(r)$ as function of
scale, normalized to an arbitrary factor for seek of clarity. The
normalization is simple because at small scales $N_c(r) = N$ where $N$
is the number of points contained in a given VL sample: in fact at
such small scales all points contribute to the statistics.  One may
note at a scale comparable but smaller than the sample size there is
an abrupt decay of this quantity: this means that only few points
contribute to the average at large scales.

That systematic fluctuations are more important than statistical ones,
can be noticed from the behavior of the conditional density in
Figs.\ref{FIGgammaR1-1}-\ref{FIGgammaR1-4} by comparing the behaviors
in the original sample (e.g. R1VL1) and in the two separate subsamples
(e.g. R1$\_$1VL1 and R1$\_$2VL1). When the distance scale approaches
the boundaries of the samples one may note that there are systematic
variations which are larger than the (small) error bars derived from
Eq.\ref{errgamma}. As already mentioned, in some cases there is an
evidence for a more flatter behavior while in other cases instead the
conditional density show a decay up to the sample boundaries which is
slower than at smaller scales. This situation puts a serious warning
for the interpretation of the large scale tail of the conditional
density. The question is how to quantify the regime where systematic
fluctuations are important and may affect the behavior of the
conditional density.

 One may define a criterion for the statistical robustness of the
 volume average, by imposing for example $N_c(r)$ to be larger than a
 certain value. While this can certainly give an useful indication, the
 problem of the volume average is more subtle. In fact when
 $r\rightarrow R_s$ there can be sufficiently enough points for
 $N_c(r)$ to be larger than a given pre-defined value: however it may
 happen that all these points lie, for example, in a cluster located
 close to the sample center. In this situation the volume average is
 not properly performed, in the sense that all points ``see'' almost
 the same volume.

A possibility to clarify such a situation has been proposed by
Joyce et al. (1999). One may compute the average distance between the
$N_c(r)$ centers at the scale $r$:
\be
R_c(r) = \frac{1}{N_c(r)(N_c(r)-1)} \sum_{i,j=1}^{N_c(r)} |\vec{r}_i -
\vec{r}_j|
\ee
where $\vec{r}_i$ and $\vec{r}_j$ are two of the $N_c(r)$ points. A
criterion for statistical validity of the volume average is then
\be
R_c\ge2\times r
\ee
which implies that the average distance between sphere centers if
larger than twice the scale at which the conditional density is
computed, assuring in this way the independence of the different terms
in the average. The values of $R_c$ for the different samples is
reported in Tab.\ref{tbl_VLSamplesProperties3} and this length-scale
is indicated as a vertical line in
Figs.\ref{FIGgamma1}-\ref{FIGgamma4}. In practice all samples show an
$R_c$ smaller than 40 Mpc/h with the exception of R1VL3 and R1VL4 for
which $R_c=54,\ 65;$ Mpc/h respectively.  Hoverer in these two samples
the conditional density does behave differently at large scales (see
Fig.\ref{gammavl3vl4}), in the sense that the change of slope 
occurs at different scales and thus at a different value average density.
\begin{figure}
\begin{center}
\includegraphics*[angle=0, width=0.5\textwidth]{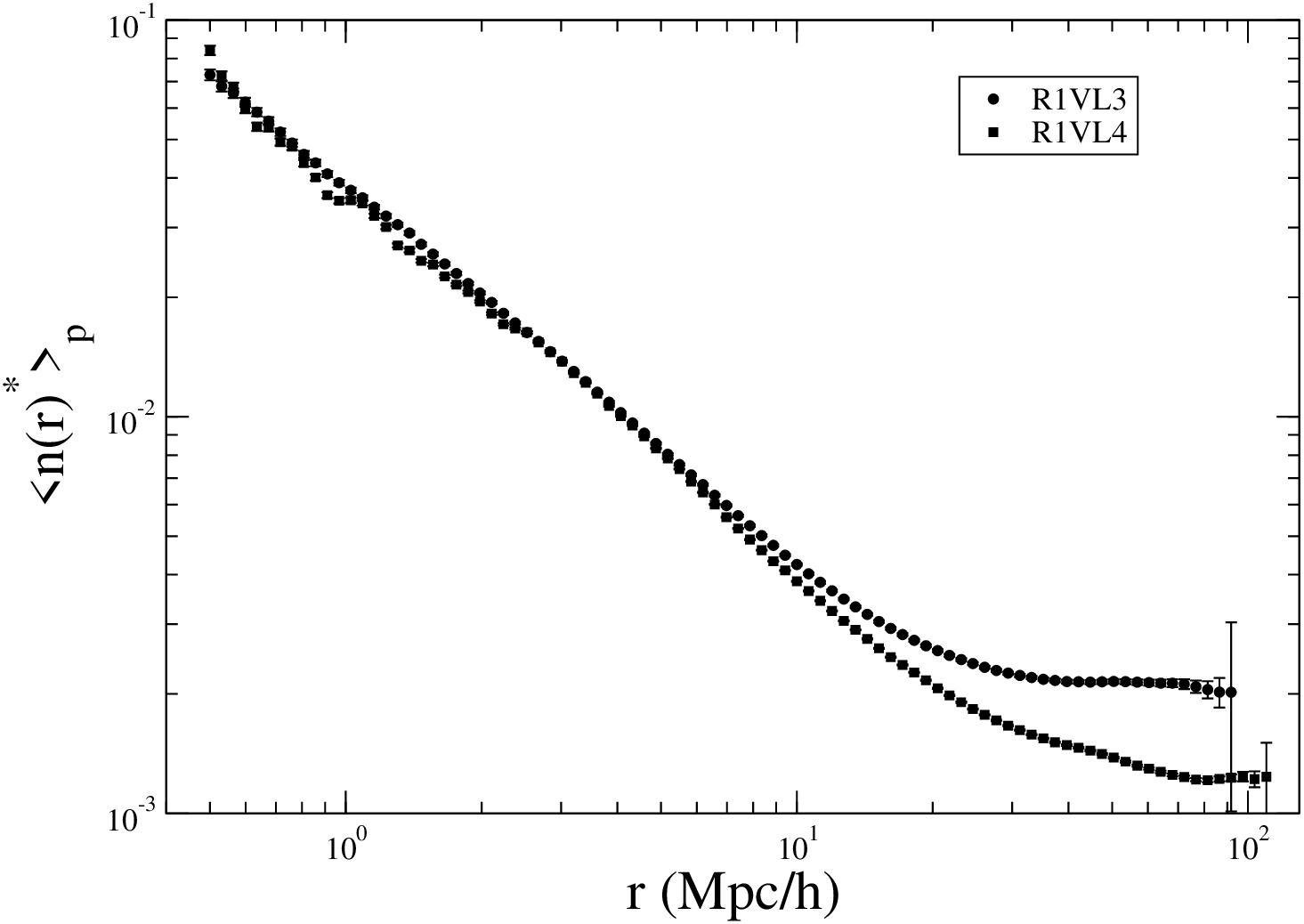}
\end{center}
\caption{Conditional density in spheres in 
the R1VL3 and R1VL4 samples, normalized to have the same amplitude at
1 Mpc/h. One may see that the large scale behavior ($r> $30 Mpc/h) is
different due to the effect of systematic fluctuations.}
\label{gammavl3vl4}
\end{figure}
Thus it is very hard to conclude about the correlation properties at
such large scales.

However we note that there is enough evidence that the signal is
smoother on scales $>$ 40 Mpc/h and that sample-to-sample fluctuations
or the variations in radial counts (discussed in Section 2) are
smaller, thus indicating a tendency toward a more uniform 
distribution.  However these data do not support unambiguously a clear
evidence in favor of homogeneity at scales of order $70$ Mpc/h, as
Hogg et al. (2005) found by analyzing the LRG sample, because the change
in correlation properties occurs at scales comparable to the scales
$R_s$ and $R_c$. We conclude that these data support an evidence for a
change of slope, with a clear tendency for $\gamma <1$, but with
undefined value.

These tests indicate that the availability of larger samples,
provided, for example, by DR5, will allow one to understand these
systematic variations. Particularly we may see that to study scales of
order 100 Mpc/h, samples with $R_s
\approx$ 300 Mpc/h are needed. However the full SDSS data will provide
us with such large and complete catalogs.


\section{Correlation properties of cosmological N-body simulations}

Gravitational clustering in the regime of strong fluctuations is
usually studied through gravitational N-body simulations.  The
particles are not meant to describe galaxies but collision-less
dark-matter mass tracers. During gravitational evolution complex
non-linear dynamics make non-linear structures at small scales, while
at large scales it occurs a linear amplification according to linear
perturbation theory. Thus, while on large scales correlation
properties do not change from the beginning --- a part a simple linear
scaling of amplitudes --- at small scales non-linear correlations are
built.  Typically in these simulations non-linear clustering is formed
up to scales of order of few Mpc.

At late times one can identify subsamples of points which trace the
high density regions, and these would  represent the
sites for galaxy formation, 
whose statistical properties are ultimately compared with
the ones found in galaxy samples. 

In order to study this problem we consider the GIF galaxy catalog
(\cite{gif}) constructed from a $\Lambda$CDM simulation run by the
Virgo consortium (\cite{virgo}).  The way in which this is done is to
firstly identify the halos, which represent almost spherical
structures with a power-law density profile from their center. The
number of galaxies belonging to each halo is set proportional to the
total number of points belonging to the halo to a certain power.  This
procedure identifies points lying in high density regions of the
dark-matter particles.  One may assign to each point a luminosity and
a color on the basis of a certain criterion which is not relevant for
what follows (see
\cite{sheth} and reference therein). 
The resulting catalog is divided into two subsamples based on
``galaxy'' color B-I as in Sheth et al. (2001): (brighter) red
galaxies (for which B-I is redder than 1.8) and (fainter) blue
galaxies (B-I bluer than 1.8).

In summary four samples of points may be considered: (i) the original
dark matter particles with $N$=$256^3$ particles (ii) all galaxies with
$N$=15445 (iii) blue galaxies with $N$=11023 and (iv) red galaxies with
$N$=4422. 

In order to understand the correlation properties in the sampled point
distributions it is useful to study the behavior of the conditional
density which, as already discussed, has a straightforward
interpretation in terms of correlations: results are shown in
Fig.\ref{gammasimu}.
\begin{figure}
\begin{center}
\includegraphics*[angle=0, width=0.5\textwidth]{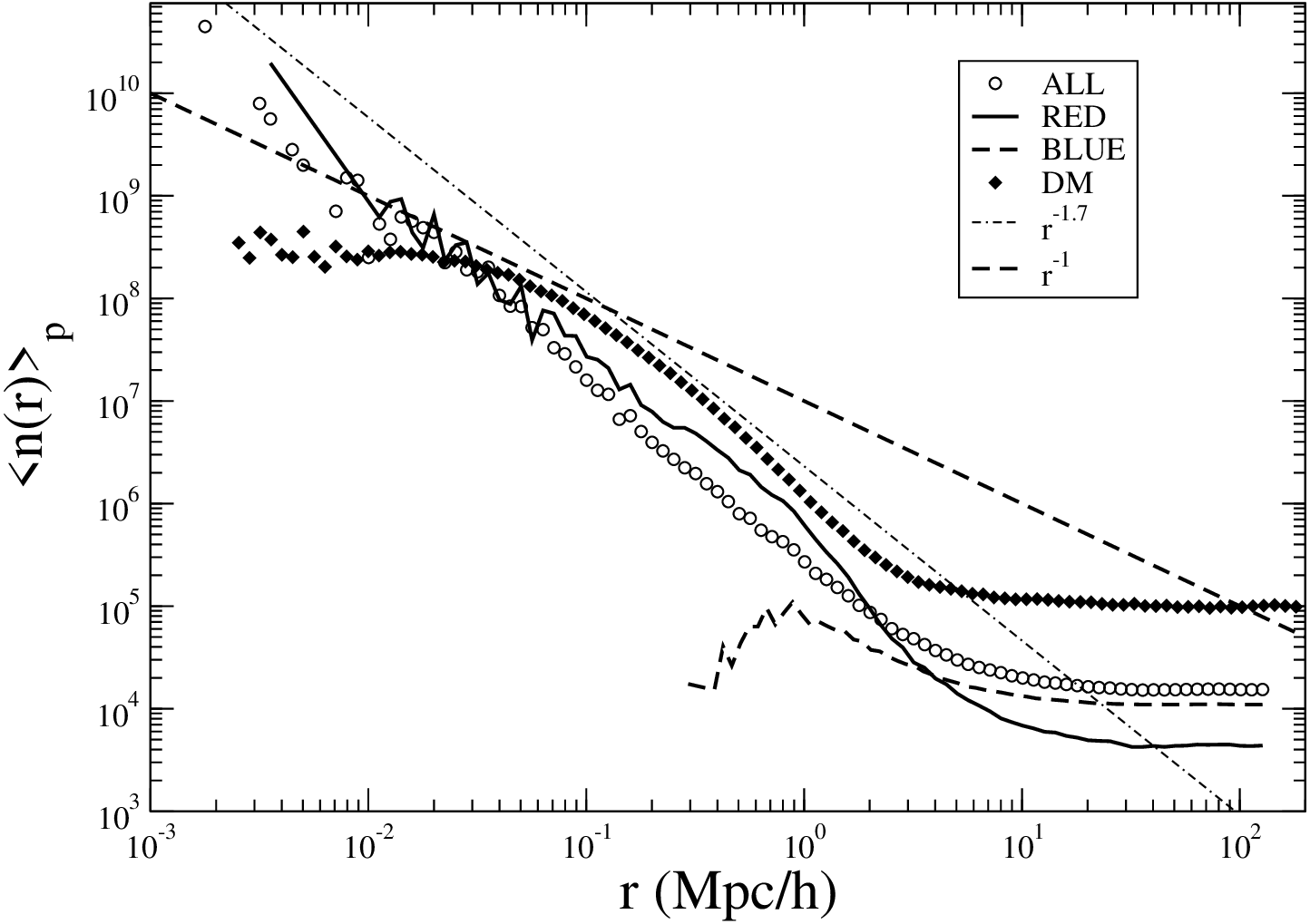}
\end{center}
\caption{Conditional density for the four samples of 
points selected in the simulation: the original dark matter (DM)
field, all ``galaxies'' (ALL), blue galaxies (BLUE) and red galaxies
(RED). The conditional density for dark matter particles (DM) has been
normalized arbitrarily. The reference dashed-dotted line has a slope
$\gamma=1.7$. The dashed line with $\gamma=1$, corresponding to the
slope measured in the galaxy samples is also reported.}
\label{gammasimu}
\end{figure}
The red galaxies are responsible for the strong correlations observed
in the full sample as the conditional density is almost the same as
for all galaxies at small scales. At large scales there is instead a
fast decrease as the sample average of red galaxies is smaller than
the one of all galaxies (there are less objects). For red galaxies the
sampling is local, i.e. their conditional density is (almost)
invariant at small scales.  Clearly, as there are globally less
objects, the sample density of red galaxies is smaller than that of
all galaxies. On the other hand blue galaxies present only some
residual correlations at small scales, and they are more numerous than
red galaxies. 

The small scale properties of these distributions can be studied
by analyzing the NN probability distribution (see Fig.\ref{FIGnnnbs}).
\begin{figure}
\begin{center}
\includegraphics*[angle=0, width=0.5\textwidth]{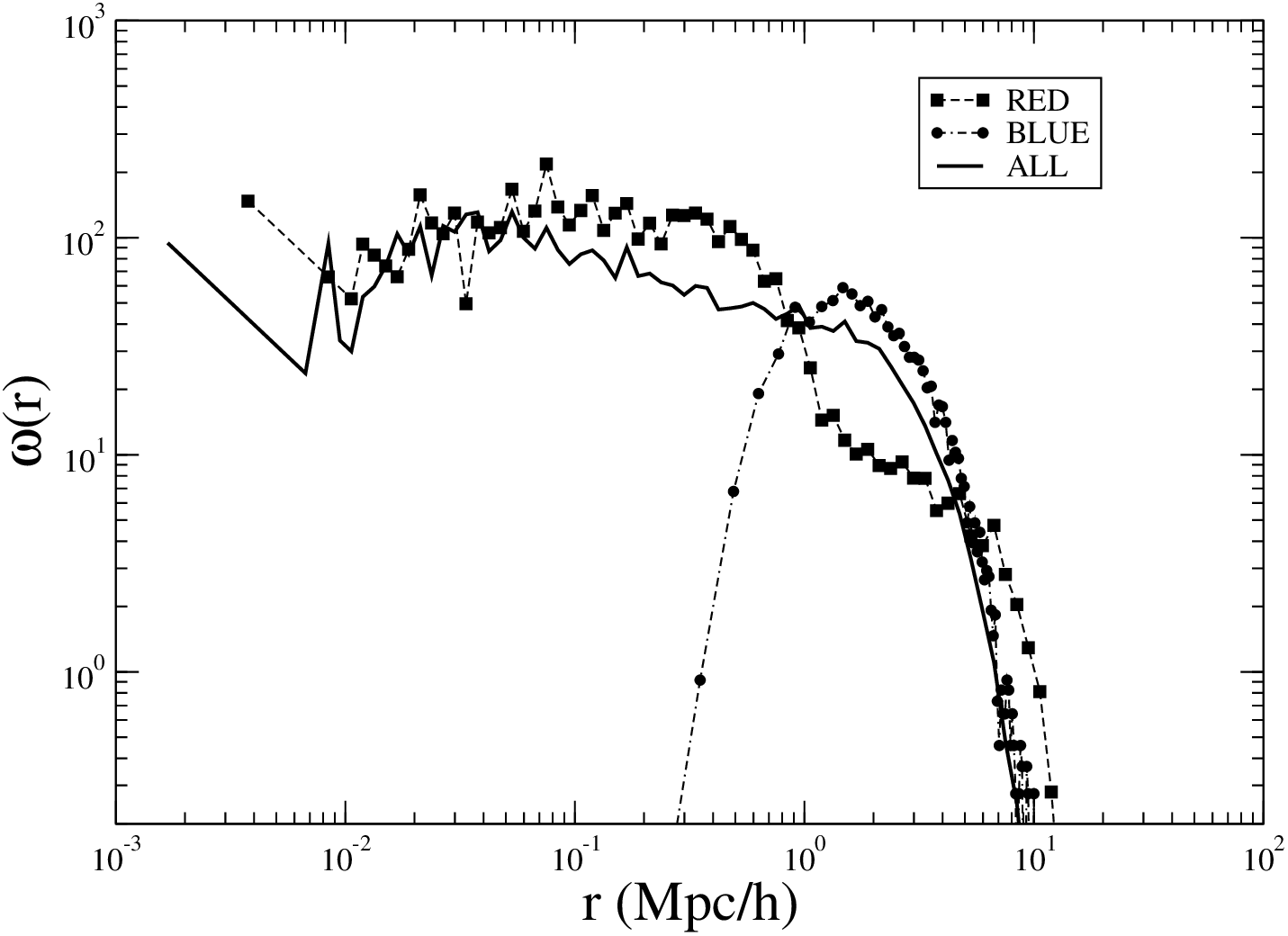}
\end{center}
\caption{Nearest neighbors probability distribution for three
point sets selected in the simulation (see discussion in the text):
all ``galaxies'' (ALL), blue galaxies (BLUE) and red galaxies (RED).}
\label{FIGnnnbs}
\end{figure}
One may note that blue galaxies have a bell-shaped distribution,
typical of the case where correlation are very weak. Instead red
and all galaxies present almost the same function, with a long
small-scale tail, which is the typical feature indicating the presence
of strong two-point correlations (see discussion in Baertschiger and
Sylos Labini, 2002). This situation is different from the one detected
in the samples of DR4 as shown in Figs.\ref{FIGnn1}-\ref{FIGnn4},
where the NN probability distribution has the same shape 
for all different samples considered.

The main points we stress are the following:

\begin{itemize} 

\item The slope of the conditional density in all the 
artificial samples considered here is different from $\gamma=1.0 \pm
0.1$ measured in the real galaxy data. In particular for those mock
samples (red galaxies, all galaxies and dark matter particles)
where correlations are power-law, the slope is $\gamma=1.7\pm 0.1$ in
the range [0.01,5] Mpc/h while a clear transition toward homogeneity
occurs at scales of order 10 Mpc/h. These different slopes
can be originated by the fact that we compare a measure in redshift space, in 
the case of real data, which can be affected by redshift distortions, with 
the mock catalogs where the conditional density has been measured in 
real space. We will examine this point in more detail in a forthcoming paper.

\item Small scales properties, as
detected by the NN probability distribution, are different in the real
and artificial samples.

\item 
The conditional densities of mock blue and red galaxies are different
at all scales and blue galaxies show almost no correlations.

\item 
Both mock red and blue galaxies show a well-defined transition to
homogeneity at a scale of oder 10 Mpc/h. As we have already mentioned, 
this is not the behavior observed in the data. Particularly the range
of non-linear structures seem to be much larger in the real data than in
the simulations.
\end{itemize} 

In conclusion, while the comparison between correlation properties of
real galaxies and mock galaxy catalogs constructed from points
selected in N-body simulations is usually performed by the analysis of
the reduced two-point correlation function, here we have presented the
comparison of the conditional density and of the NN probability
distributions. We find that some important disagreement between data
and simulations are evident when the behavior of these statistical
quantities are considered. This is not the same conclusion that one
may reach by analyzing the reduced correlation function $\xi(r)$: the
reason is that in the estimation of $\xi(r)$ one uses the estimation
of the sample average which introduces a finite-size effect which may
affect both the amplitude and slope of this function (see
e.g. Gabrielli et al., 2004 for a detailed discussion of this
point). The estimation of the conditional density is less affected by
finite-volume effects and the comparison between different sample is
straightforward.

 Note that the data are analyzed in redshift space and the simulations
 in real space. However given that velocities are typically smaller
 than $500$ km/s the difference between real and redshift space cannot
 be accounted by the effects of peculiar velocities on scales larger
 than 5 Mpc/h. The problem of the relation between real and redshift
 space, considering the finite size effects present when strong
 correlation characterize the data, has been discussed in Vasilyev,
 Baryshev \& Sylos Labini (2006).


\section{Discussion and Conclusions}

Our main results are the following:

(i) In all VL samples we find that in the range of scales $0.5 \le r
\ltapprox 30$ Mpc/h the conditional density shows power-law
correlation with a power-law index $\gamma =1.0\pm 0.1$. This result
is in good agreement with the behavior found in other smaller samples
by Sylos Labini et al. (1998), Joyce et al. (1999) and in the SDSS LRG
sample by Hogg et al. (2005), and with the correlation properties
measured by Vasilyev, Baryhsev \& Sylos Labini (2006) in the
2dFGRS. 

Note that we do not confirm the results of Zehavi et al. (2004) who
found a departure from a power-law in the galaxy correlation function
at a scale of order 1 Mpc/h: their analysis has been performed in real
space while ours is in redshift space. In this range of scales
nearest-neighbor correlation dominate the behavior of the conditional
density and thus also of the reduced correlation function and for a
detailed understanding of this regime a study of the nearest-neighbor
is shown to be necessary.

In addition we do not find either a luminosity or color dependence of
the galaxy the conditional density in the regime where the statistics
is robust. In this respect Zehavi et al. (2005) have considered the
behavior of the reduced two-point correlation function, and concluded
that there is a color (luminosity) dependence of galaxy
correlations. This apparent disagreement can be understood by
considering that the reduced two-point correlation function can be
strongly affected by finite-size effects in the regime where the
conditional density presents power-law correlations (see discussion,
e.g. in Joyce et al., 2005).  Moreover results by Zehavi et al. (2005)
have been obtained in real space: in Vasilyev, Baryhsev
\& Sylos Labini (2006) we discussed the kind of finite size effects
which perturb the estimation of $\xi(r)$ when the conditional density
has power-law correlations.

(ii) In the range $30 \ltapprox r \ltapprox 100$ Mpc/h the situation
is less clear: as we discussed finite volume effects are important in
this range of scales and systematic unaveraged fluctuations may affect
the results. We have presented several tests to show the role of
finite volume effects and to determine the range of scales where they
perturb the estimation of the conditional density, finding that in all
but two samples the volume average is properly performed up to $R_c
\approx 40$ Mpc/h. In the remaining two samples we have shown that
systematic fluctuations persist up to their boundaries $R_s$.

Thus in the range $30 \ltapprox r \ltapprox 100$ Mpc/h we find
evidences for more uniform distribution and hence a smaller power law
index ($\gamma <1$) in the conditional density. This is a stable
result in all samples considered. However a detailed analysis of the
behavior of the conditional density in all samples does not allow us
to conclude neither that there is definitive crossover to homogeneity
at a scales of order 70 Mpc/h as Hogg et al. (2005) have concluded by
considering the LRG sample, nor that there is a change of power-law
index beyond 30 Mpc/h which remain stable up to samples limit, i.e. up
to 100 Mpc/h. Both possibilities are still open and will be clarified
by forthcoming data releases of SDSS as the solid angle is going to
sensibly grow.

(iii) The comparison of mock galaxy catalogs constructed from
particle distributions extracted from cosmological N-body simulations
with real galaxy data outlines a problematic situation.  From the one
hand we have discussed the fact that the slope of the conditional
density in latter case is different from the one measured in real
catalogs.  On the other hand we have also stressed that when
constructing artificial galaxy samples from dark matter particles in
N-body simulations, there are different behaviors in the conditional
density according to the different selection criteria used, and thus
on the different way to assign ``luminosity'' and ``color'' to the
artificial galaxies. In any case, this behavior is not in agreement
with the data, as in all samples here analyzed, the same slope in the
conditional density is measured.  The same situation is present when
the NN probability distribution is considered.  Then in N-body
simulations structures are sensibly smaller than in real data, as
shown by the definitive crossover to homogeneity at about 10 Mpc/h
found in N-body particle distribution, contrary to the galaxy case
where the crossover may happen on much larger scales of order 100
Mpc/h.

It is worth noticing that we have used a very conservative statistical
analysis which introduces an important constraints on the way we treat
the data.  For example if the distribution would have been uniform on
scales smaller than the actual sample sizes, the conditional density
estimation can be done for all points in the sample, even on large
scales, not just the points near the center of the sample, because it
can be assumed that volume outside the survey region is statistically
similar to volume inside. This is the standard approach with
conventional two-point statistics in the literature. On the other hand
we have used, for example, periodic boundary conditions in the
analysis of artificial simulations, as in this case the distribution
is periodic, beyond the simulation box, by construction. However, as
we do not know whether this is case for galaxy distribution, and
actually we would like to test this point, we have used more
conservative statistics to analyze the real data. This, instead of
being a limitation, allow us to derive results about galaxy
correlation properties which are unbiased by finite size effects.
Indeed, when using less conservative methods, one is implicitly making
the assumption that finite size effects, induced by long range
correlations in the galaxy distribution, are negligible. Here we
instead test that this is the case in the data we consider and
actually we find evidence that, because of the long range nature of
galaxy correlations, there are subtle finite size effects which should
then put a serious warning on the use of less conservative statistical
methods. Having used a more conservative statistics we are able to
obtain results which are less biased by finite size effects (which
ultimately appear from the presence of large fluctuations represented
by large scale structures) with respect to the ones derived by a
statistical analysis which makes use of some untested assumptions to
derive its results. For example we get that the exponent of the
conditional density is -1 instead of -1.7 as derived through a more
``relaxed'' analysis, at the same scales. The measurements of the
conditional density has been performed in real space in the mock
catalogs and in redshift space in the real samples, and this can be
the origin of the different values of the correlation exponents.
Whether this is case, or a finite  size effect is playing a crucial role
will be studied in a forthcoming paper.

Finally we would like to briefly discuss our results in relation to
theoretical models of fluctuations in standard cosmologies.  It has
been shown (see e.g. Gabrielli et al. 2004) that the only feature of
the primordial correlations, defined in theoretical models like the
cold dark matter (CDM) one, which can be detected in galaxy data is
represented by the large scale tail of the reduced correlation
function. In fact, in terms of correlation function $\xi(r)$ CDM
models presents the following behavior: it is positive at small
scales, it crosses zero at a certain scale and then it is negative
approaching zero with a tail which goes as $r^{-4}$ in the region
corresponding to $P(k) \sim k$ (see e.g. Gabrielli et al. 2004). The
super-homogeneity (or Harrison-Zeldovich) condition says that the
volume integral over all space of the correlation function is zero
\be
\int_0^{\infty} d^3r \xi(r) = 0 \;.
\ee
This means that there is a fine tuned balance between small-scale
positive correlations and large-scale negative anti-correlations. This
is the behavior that one would like to detect in the data in order to
confirm inflationary models. Up to now this search has been done
through the analysis of the galaxy power spectrum (PS) which should
scale as $P(k) \sim k$ at small $k$ (large scales). No observational
test of this behavior has been provided yet.  However for this case
one should consider an additional complication.

In standard models of structure formation galaxies result from a {\it
sampling} of the underlying CDM density field: for instance one
selects only the highest fluctuations of the field which would
represent the locations where galaxy will eventually form. It has been
shown that sampling a super-homogeneous fluctuation field changes the
nature of correlations (Durrer et al., 2003). The reason for this can
be found in the property of super-homogeneity of such a distribution:
the sampling necessarily destroys the surface nature of the
fluctuations, as it introduces a volume (Poisson-like) term in the
mass fluctuations, giving rise to a Poisson-like PS on large scales
$P(k)\sim$ constant.  The ``primordial'' form of the PS is thus not
apparent in that which one would expect to measure from objects
selected in this way. This conclusion should hold for any generic
model of bias and its quantitative importance has to be established in
any given model (Durrer et al., 2003).

On the other hand one may show (Durrer et al., 2003) that the negative
$r^{-4}$ tail in the correlation function does not change under
sampling: on large enough scales, where in these models (anti)
correlations are small enough, the biased fluctuation field has a
correlation function which is linearly amplified with respect to the
underlying dark matter correlation function. For this reason the
detection of such a negative tail would be the main confirmation of
models of primordial density field. This will be possible if firstly a
clear determination of the homogeneity scale will be obtained, and
then if the data will be statistically robust enough to allow the
determination of the correlation when it is $\xi(r) \ll 1$. While
Eiseinstein et al. (2005) claimed to have measured that $\xi(r)
\approx 0.01$ at scales of order 100 Mpc/h in a sample of SDSS LRG
galaxies, here we cannot confirm these results as our analysis does
not extend to such large scales with a robust statistics. However
from the large fluctuations observed, for example in the behavior of
the radial counts and in sample-to-sample variations of the
conditional density at such large scales, we conclude that this result
deserves more studies, and perhaps much larger samples, to be
confirmed.


\section*{Acknowledgments}

We thank Andrea Gabrielli, Michael Joyce and Luciano Pietronero for
useful discussions and comments.  Yu.V.B. and N.L.V. thank the
``Istituto dei Sistemi Complessi'' (CNR, Rome, Italy) for the kind
hospitality during the writing of this paper.  FSL acknowledge the
financial support of the EC grant No. 517588 ``Statistical Physics for
Cosmic Structures" and the MIUR-PRIN05 project on ``Dynamics and
Thermodynamics of systems with long range interactions" for financial
supports. Yu.B and N.V thanks the partial financial support by Russian
Federation grants NSh-8542.2006.2 and RNP.2.1.1.2852.


{}

\end{document}